\pdfoutput=1
\documentclass[12pt,onecolumn,draftclsnofoot]{IEEEtran}

\usepackage{amsmath,amsfonts,amssymb}
\usepackage{amsthm}
\usepackage{algorithm}
\usepackage{algpseudocode}
\usepackage{array}
\usepackage{booktabs}
\usepackage{multirow}
\usepackage{multicol}
\usepackage{arydshln}
\usepackage{enumitem}
\usepackage{enumerate}
\usepackage[caption=false,font=normalsize,labelfont=sf,textfont=sf]{subfig}
\usepackage{textcomp}
\usepackage{stfloats}
\usepackage{url}
\usepackage{graphicx}
\usepackage{epstopdf}
\usepackage{bm}
\usepackage{dsfont}
\usepackage{xcolor}
\usepackage[numbers,sort&compress]{natbib}
\usepackage[unicode=true,bookmarks=false]{hyperref}
\allowdisplaybreaks[4]
\graphicspath{{Figs/}}

\newcommand{\vA}{\boldsymbol{A}}

\newcommand{\vC}{\boldsymbol{C}}
\newcommand{\vD}{\boldsymbol{D}}

\newcommand{\vI}{\boldsymbol{I}}

\newcommand{\vO}{\boldsymbol{O}}

\newcommand{\vQ}{\boldsymbol{Q}}

\newcommand{\vS}{\boldsymbol{S}}

\newcommand{\vU}{\boldsymbol{U}}
\newcommand{\vV}{\boldsymbol{V}}

\newcommand{\vY}{\boldsymbol{Y}}

\newcommand{\vb}{\boldsymbol{b}}

\newcommand{\ve}{\boldsymbol{e}}

\newcommand{\vg}{\boldsymbol{g}}
\newcommand{\vh}{\boldsymbol{h}}

\newcommand{\vr}{\boldsymbol{r}}

\newcommand{\vu}{\boldsymbol{u}}

\newcommand{\vx}{\boldsymbol{x}}
\newcommand{\vy}{\boldsymbol{y}}

\newcommand{\bLambda}{\boldsymbol{\Lambda}}

\newcommand{\bSigma}{\boldsymbol{\Sigma}}

\newcommand{\bxi}{\boldsymbol{\xi}}
\newcommand{\btheta}{\boldsymbol{\theta}}
\newcommand{\bphi}{\boldsymbol{\phi}}
\newcommand{\bPhi}{\boldsymbol{\Phi}}

\newcommand{\bOmega}{\boldsymbol{\Omega}}

\newcommand{\Ex}{\mathbb{E}}

\newtheorem{definition}{Definition}
\newtheorem{theorem}{Theorem}
\newtheorem{lemma}{Lemma}

\newtheorem{corollary}{Corollary}
\newtheorem{assumption}{Assumption}
\newtheorem{remark}{Remark}
\newtheorem{appendixlemma}{Lemma}

\begin{document}
\title{FAST: Functional Adaptive Sensing and Testing for Quickest Change Detection}

\author{Peiyao Liu, Jie Guo, Zhijun Wang, Xueru Zhang, and Nan Chen%
\thanks{P. Liu and N. Chen are with the Department of Industrial Systems Engineering and Management, National University of Singapore, Singapore (e-mail: liupy25@nus.edu.sg; isecn@nus.edu.sg). N. Chen is the corresponding author.}%
\thanks{J. Guo is with the College of Economics and Management, Nanjing University of Aeronautics and Astronautics, Nanjing, China (e-mail: guojie25@nuaa.edu.cn).}%
\thanks{Z. Wang is with the Department of Industrial Engineering and Management, Shanghai Jiao Tong University, Shanghai, China (e-mail: ZJ-Wang@sjtu.edu.cn).}%
\thanks{X. Zhang is with the Department of Statistics, University of Science and Technology Beijing, Beijing, China (e-mail: xueruzhang@ustb.edu.cn).}}

\maketitle

\begin{abstract}
Quickest change detection (QCD) is studied for partially observed functional data streams. At each sampling epoch, a function is realized over a continuous domain, but its values can be observed at only a fixed number of locations. At an unknown change point, the functional mean undergoes a shift that is assumed to admit a sparse representation in a localized basis. We propose Functional Adaptive Sensing and Testing (FAST), which jointly estimates the unknown post-change parameter, selects sensing locations, and performs sequential detection. A minimax lower bound on the worst-case detection delay is derived under an average-run-length constraint. A non-asymptotic estimation-error bound is established for the policy-dependent observations induced by adaptive sensing. Under the stated regularity conditions, FAST is first-order asymptotically optimal when sensing subsets are optimized exactly. Simulations and a wildfire case study demonstrate the performance of the proposed procedure.
\end{abstract}

\begin{IEEEkeywords}
Quickest change detection, adaptive sensing, functional data, partial observation, sparse estimation
\end{IEEEkeywords}

%----------------------------------------------------------------------------------------- introduction ---------------------------------------------------------------------------------------------------------------
\section{Introduction}

Quickest change detection (QCD) concerns detecting a change in the distribution of sequential observations as rapidly as possible while controlling false alarms \citep{page1954continuous,lorden1971procedures}. In classical formulations, the observation at each time is fully available, and the detector only needs to decide when to stop and declare a change. In many modern sensing systems, however, acquiring a complete high-dimensional or function-valued observation at every time may be infeasible because of sensing, communication, or computational constraints. The detector must therefore decide not only when to stop, but also what to observe.

We focus on \textit{functional data streams}, represented by a sequence $\{y_t(s):t\in\mathbb N\}$, where at each time $t$, $y_t(s)$ is a function defined over a continuous domain $\mathcal S$. Unlike finite-dimensional multivariate vectors, functional data are inherently infinite-dimensional and are typically recorded on dense or irregular grids \citep{fda}. Examples arise in yield curves \citep{shang2017grouped}, sales trajectories \citep{lei2023new}, functional magnetic resonance imaging \citep{qiao2019functional}, and continuous PM$_{2.5}$ profiles \citep{wen2023ℓ0}. Such observations can contain rich information about structural changes, but acquiring the complete function at every time point may be impractical. In particular, energy limitations in continuous sensing systems, communication constraints in limited-bandwidth networks, and the computational cost of real-time processing may limit sensing to a small subset of locations at each time step.

Under this sensing constraint, the detector must identify a structural change as quickly as possible while simultaneously deciding where to sample the function. The sensing policy determines the statistical information available to the detector: concentrating observations on locations that appear informative can accelerate evidence accumulation, but insufficient exploration can leave a localized change undetected. The resulting problem therefore involves a fundamental exploration--exploitation tradeoff between learning the unknown post-change structure and collecting information for rapid detection.

We address this tradeoff through adaptive sensing. At each time $t$, the detector selects a subset $Z_t\subset\mathcal S$ and observes the function only at the selected locations. Since observations at unselected locations are unavailable, the sensing decision has a partial-feedback structure. We formulate the resulting subset-selection problem through a combinatorial multi-armed bandit (CMAB) mechanism, using Kullback--Leibler (KL) information to quantify the contribution of a sensing subset to sequential detection. The post-change shift is assumed to be localized and is represented as a sparse shift in a localized basis expansion. This structure enables sparse online estimation while the sensing policy adapts to the information accumulated over time.

Existing limited-sensing change-detection methods primarily consider finite-dimensional vectors, images, or tensors \citep{guo2023bayesian,gomez2022adaptive,yao2024adaptive}. In contrast, location-level partial observation of a functional domain introduces both an infinite-dimensional representation problem and a sequential sensing problem. At the same time, existing high-dimensional bandit methods generally study fixed-parameter learning in stationary environments and evaluate performance through cumulative regret \citep{bastani2020online,wang2023efficient}. In QCD, the parameter changes at an unknown time, and performance is instead governed by detection delay under a false-alarm constraint. A detailed review of the relevant QCD, limited-sensing, and high-dimensional bandit literature is provided in Section \ref{sec:lit}.

To address these challenges, we develop FAST (Functional Adaptive Sensing and Testing), a policy for QCD in sparsely sampled functional data. The proposed framework addresses three main challenges: (1) representing localized structural changes over a continuous functional domain; (2) estimating an unknown sparse post-change shift from adaptively sampled observations while limiting contamination from pre-change data; and (3) selecting a restricted subset of sensing locations so that detection information is accumulated efficiently under estimation uncertainty. We use a localized basis expansion to obtain a sparse coefficient representation, a sliding-window Lasso estimator to learn the post-change parameter, a KL-information-based UCB rule to select sensing locations, and a recursive likelihood-ratio statistic for sequential detection. Our main contributions are as follows:
\begin{itemize}
\item We formulate QCD for sparsely sampled functional data streams, where each function is observed at only a limited number of locations. A localized basis representation converts localized functional mean shifts into sparse coefficient shifts, allowing the unknown post-change parameter to be learned jointly with the sensing policy.

\item We characterize the fundamental detection limit by deriving a minimax lower bound on the worst-case average detection delay. We further establish a non-asymptotic error bound for the online Lasso estimator under the policy-dependent, non-i.i.d. observations generated by adaptive sensing. Using these results, we show that FAST is first-order asymptotically
optimal under exact subset optimization.

\item We establish false-alarm control and characterize the information collected by the adaptive sensing rule. For implementation, a greedy subset-selection rule reduces the sensing optimization from exhaustive combinatorial search to $O(mM)$ objective evaluations, while retaining the approximation factor appearing in the detection-delay guarantee.
\end{itemize}

The remainder of this paper is organized as follows. Section \ref{sec:lit} reviews related work in quickest change detection, change detection with limited sensing, and high-dimensional bandits. Section \ref{sec:problem_formulation} formalizes the QCD problem. Section \ref{sec:method} details the proposed FAST methodology. Section \ref{sec:theory} presents the theoretical properties of the framework. Section \ref{sec:numerical} evaluates the numerical performance, followed by a real-world case study. Finally, Section \ref{sec:conclusion} concludes the paper.

\section{Related Work}
\label{sec:lit}

\subsection{Quickest Change Detection}

Classical QCD assumes that sequential observations are fully available to the detector. Under i.i.d. pre- and post-change models, the CuSum procedure of \citet{page1954continuous} and Lorden's minimax criterion \citep{lorden1971procedures} form a central foundation of the theory. \citet{lai1998information} developed information-theoretic lower bounds and nearly optimal detection rules for general stochastic systems with parameter changes. The theory was subsequently extended to more general observation models. In particular, \citet{tartakovsky2017asymptotic} established asymptotic optimality results for sequential changepoint detection with non-i.i.d. observations, while \citet{tartakovsky2019mixture} studied mixture rules for general stochastic models with uncertainty in the post-change distribution. More recently, \citet{liang2023quickest} considered non-stationary post-change observations and developed window-limited CuSum and GLR-CuSum procedures that attain a universal asymptotic lower bound under known and parametrically uncertain post-change models, respectively.

These works provide important theoretical foundations for QCD but generally assume that the observations used by the detector are fully available at each sampling epoch. Our setting instead imposes a within-function sensing constraint: only a small subset of locations on each functional observation can be acquired, and the sensing subset itself must be selected sequentially. This creates a coupled sensing-and-detection problem, which is reviewed next.

\subsection{Limited-Sensing Change Detection}

Limited-sensing change detection studies how to detect distributional changes when only part of an underlying observation can be acquired at each decision epoch. Existing methods have primarily focused on finite-dimensional multivariate vectors, in which a subset of coordinates or sensor channels is selected for observation, and on structured data such as images and tensors. A common approach is to cast the sensing decision as a multi-armed or combinatorial bandit problem, creating an exploration--exploitation tradeoff between collecting information from currently suspicious components and continuing to explore other components. UCB and Thompson sampling (TS) provide two representative mechanisms for this tradeoff \citep{lai1985asymptotically,auer2002finite,rusmevichientong2010linearly,abbasi2011improved,thompson1933likelihood,agrawal2013thompson,russo2016information,russo2018learning}.

Bandit-based sensing rules have consequently been combined with sequential detection statistics in several forms. For example, \citet{zhang2019partially} combined an exponentially weighted moving average-type statistic with a UCB sensing policy and constructed a coordinate-wise confidence bound for a nonlinear super-arm reward. \citet{zhang2023bandit} adopted a Shiryaev--Roberts-type statistic with TS, and \citet{guo2023bayesian} further incorporated sparse change structures. Beyond bandit-based sensing, imputation-then-detection methods first reconstruct unobserved components and then apply a sequential detection statistic to the augmented data \citep{xian2021online,zan2023spatial}. Active-sensing approaches have also been developed for multichannel settings. For example, \citet{chaudhuri2021sequential,xu2023asymptotic} proposed policies that continue observing one variable until it is deemed unchanged or a sampling threshold is reached before moving to another variable, while \citet{tsopelakos2022sequential} allowed the set of variables available for sensing to vary over time and proposed a probabilistic sensing rule.

Most of this literature is formulated for discrete multivariate vectors. More recent work has considered images or tensors \citep{gomez2022adaptive,yao2024adaptive}, but location-level partial observation within a functional domain remains largely unexplored. \citet{yao2023adaptive} considered multivariate functional observations $\vy_t(s)=[y_{t1}(s),\ldots,y_{tp}(s)]$, but selecting channel $j$ reveals the entire function $y_{tj}(s)$ over $s\in\mathcal S$. Hence, their sensing decision is made at the channel level, $j\in\{1,\ldots,p\}$, rather than at the location level, $Z_t\subset\mathcal S$. The present problem instead requires selecting only a small number of locations within each functional observation. This distinction motivates a sensing-and-detection framework specifically for sparsely sampled functional data.

\subsection{High-Dimensional Bandits}

In sequential detection, post-change shifts are often sparse in functional representations. This sparsity naturally connects our problem to high-dimensional bandits, where sparse estimators are integrated into online decision-making to overcome the curse of dimensionality. Early studies explored different forms of regularization, such as the minimax concave penalty \citep{wang2018minimax} and doubly robust estimation \citep{kim2019doubly}. As a theoretical cornerstone, \citet{bastani2020online} established a Lasso bandit framework with forced exploration and proved the convergence of the Lasso estimator. Subsequent work further relaxed the reliance on prior knowledge of the sparsity structure \citep{oh2021sparsity, ariu2022thresholded}.

To account for parameter estimation uncertainty, another line of work incorporates confidence-bound mechanisms into high-dimensional bandits. \citet{abbasi2012online} developed online-to-confidence-set conversions, enabling optimistic exploration in sparse settings. More recently, \citet{wang2023efficient} combined random projection with Lasso-UCB to construct confidence ellipsoids efficiently in high-dimensional spaces. Despite these advances, existing high-dimensional bandit methods primarily address fixed-parameter learning, leaving the sparsity-aware quickest detection of unknown parameter changes largely unexplored.

%However, a critical limitation of these approaches is their predominant reliance on forced random sampling or greedy exploitation, which does not explicitly account for parameter estimation uncertainty.

%%%%%%%%%%%%%%%%%%%%%%%%%%%%%%%%%%%%%%%%%%%%%%%%%%%%%%%%%%%%%%%%%%%%%%%%%%
\section{Problem Setup}
\label{sec:problem_formulation}
In this section, we first introduce the QCD formulation, and then derive a minimax lower bound for our problem.
%Specifically, $\btheta$ is an exact finite-dimensional representation of the mean function, whose definition is deferred to Section \ref{sec:method}. 

\subsection{Quickest Change Detection (QCD)}
We formulate a QCD problem for sparsely sampled functional data streams $\{y_t(s): t \in \mathbb{N}\}$ defined over a continuous domain $\mathcal{S}$, which can flexibly represent a 1D interval, a 2D surface, or a 3D space. We assume the functional data $y_t(s)$ are independent across time $t$, with their distribution governed by a parameter $\btheta$. Let $\tau_0 \in \mathbb{N}$ denote the unknown change-point. The system is in a pre-change regime with $\btheta = \btheta_0$ when $t< \tau_0$, and shifts to a post-change regime with $\btheta = \btheta_1$ when $t \ge \tau_0$. Without loss of generality, we set $\btheta_0 = \mathbf{0}$ and $\btheta_1 \neq \mathbf{0}$. Due to limited sensing resources, the full functional data cannot be observed at each time step. Instead, the learner chooses a sensing action $Z_t \subset \mathcal{S}$ to observe $m$ discrete points such that $|Z_t| = m$, obtaining the subsampled observation $\vY_{Z_t} = \{y_t(s) : s \in Z_t\}$. We denote $\mathcal{F}_t = \sigma(Z_1, \vY_{Z_1}, \dots, Z_t, \vY_{Z_t})$ as the natural filtration representing the causal history of sensing actions and subsampled observations up to time $t$, with $\mathcal{F}_0$ being the trivial $\sigma$-algebra. At each round $t$, relying exclusively on the available history $\mathcal{F}_{t-1}$, the learner decides either (1) to continue sensing by selecting a new sensing action $Z_t$, or (2) to stop and declare that a change has occurred.

Formally, a general detection scheme is characterized by a pair $(Z, \tau_1)$. Here, $Z = \{Z_t\}_{t \ge 1}$ is a dynamic sensing policy where each action $Z_t$ is strictly $\mathcal{F}_{t-1}$-measurable (i.e., non-anticipating), and $\tau_1$ is an $\{\mathcal{F}_t\}$-adapted stopping time indicating the declaration of a change. Typically, the stopping time takes the form $\tau_1 = \inf\{t \ge 1 : W_t > b\}$, where $W_t$ is a recursively updated detection statistic and $b$ is a predefined threshold. 
An ideal detection scheme must balance two competing objectives: controlling false alarms in the pre-change regime and minimizing detection delay after the change occurs. The false-alarm performance is governed by the average run length ($\text{ARL}_0$), defined as $\text{ARL}_0 = \Ex_{\infty}[\tau_1]$, where $\Ex_{\infty}$ denotes the expectation under the pre-change regime (i.e., $\tau_0 = \infty$). In practice, a detection scheme must ensure that its $\text{ARL}_0$ is bounded below by a prescribed constant $\gamma > 1$. We formalize this constraint by defining the class of admissible policies. 

\begin{definition}[Admissible Policy Class]
For a prescribed constant $\gamma > 1$, we define the admissible policy class $\mathcal A_\gamma$ as the set of all detection schemes that satisfy $\mathrm{ARL}_0\ge \gamma$:
\begin{equation*}
    \mathcal{A}_{\gamma} = \{ (Z, \tau_1) : \Ex_{\infty}[\tau_1] \ge \gamma \}.
\end{equation*}
\end{definition}

As a measure of the responsiveness of the detection schemes, the standard average detection delay (ADD) is typically expressed as $\Ex_{\tau_0, \btheta_1}[(\tau_1 - \tau_0 + 1)^+]$, where $\Ex_{\tau_0, \btheta_1}$ denotes the expectation when the true change occurs at time $\tau_0$. However, since $\tau_0$ is unknown, and limited sensing makes the available information heavily dependent on the historical sensing path $\mathcal{F}_{\tau_0-1}$, we adopt a worst-case measure for detection delay (WADD) based on the widely used Lorden's minimax criterion \citep{lorden1971procedures}.

\begin{definition}[Worst-Case Average Detection Delay]
For a given post-change parameter $\btheta_1$, the WADD is defined as the supremum over all possible change-points $\tau_0 \ge 1$ and all possible history paths $\mathcal{F}_{\tau_0-1}$ prior to the change:
\begin{equation*}
    J_{\btheta_1}(Z, \tau_1) = \sup_{\tau_0 \ge 1} \operatorname{ess\,sup} \Ex_{\tau_0, \btheta_1}[(\tau_1 - \tau_0 + 1)^+ \mid \mathcal{F}_{\tau_0-1}].
\end{equation*}
\end{definition}

Consequently, the QCD problem is formulated as a constrained optimization problem: finding an admissible policy $(Z, \tau_1) \in \mathcal{A}_{\gamma}$ that minimizes the WADD $J_{\btheta_1}(Z, \tau_1)$. In the full-data setting, classical CuSum procedures provide well-established QCD guarantees. Under adaptive sensing, however, the problem becomes more challenging because the sensing locations must also be selected online to minimize the detection delay.

\subsection{Minimax Lower Bound}
To establish the fundamental limits of the QCD problem, we derive a minimax lower bound on the WADD. Intuitively, a change is easier to detect when the post-change distribution is more distinguishable from the pre-change distribution. This statistical separation is measured by the KL information \citep{gopalan2021bandit,veeravalli2024quickest}. Therefore, the benchmark delay is governed by the maximum KL information that an oracle sensing policy can collect per observation, as formalized in Definition \ref{def:max_info_gain}.

\begin{definition}[Oracle KL Information]
\label{def:max_info_gain}
Let $\mathbb{P}_{Z, \btheta_1}$ and $\mathbb{P}_{Z, \mathbf{0}}$ be the probability measures of the subsampled observation $\vY_{Z}$ under the post-change parameter $\btheta_1$ and the pre-change parameter $\mathbf{0}$, respectively. For any candidate sensing subset $Z$, we define the instantaneous KL information as
\begin{equation*}
\label{eq:info}
    \mathcal{I}(Z, \btheta_1) := D_{\text{KL}}(\mathbb{P}_{Z, \btheta_1} \parallel \mathbb{P}_{Z, \mathbf{0}}).
\end{equation*}

The oracle KL information associated with the optimal sensing subset is then defined as
\begin{equation*}
\label{eq:reward}
    \mathcal{I}^*(\btheta_1) := \max_{Z: |Z|=m} \mathcal{I}(Z, \btheta_1).
\end{equation*}
\end{definition}

Based on these definitions, Theorem \ref{thm:lower_bound} presents the minimax asymptotic lower bound for the WADD. The proof is provided in Appendix~B.1. This result shows that, under the $\text{ARL}_0$ constraint $\gamma$, the best achievable WADD scales inversely with the oracle KL information $\mathcal{I}^*(\btheta_1)$. Our objective is therefore to design an admissible sensing-and-stopping policy whose WADD approaches this fundamental limit.

\begin{theorem}[Minimax Lower Bound]
\label{thm:lower_bound}
For any admissible policy $(Z, \tau_1) \in \mathcal{A}_{\gamma}$ and any post-change parameter $\btheta_1$ with $0<\mathcal I^*(\btheta_1)<\infty$, as $\gamma \to \infty$, the WADD satisfies:
\begin{equation*}
    \inf_{(Z, \tau_1) \in \mathcal{A}_{\gamma}} J_{\btheta_1}(Z, \tau_1) \ge \frac{\log \gamma}{\mathcal{I}^*(\btheta_1)} (1 + o(1)).
\end{equation*}
\end{theorem}

%%%%%%%%%%%%%%%%%%%%%%%%%%%%%%%%%%%%%%%%%%%%%%%%%%%%%%%%%%%%%%%%%%%%%%%%%%%%%%%%%%%%%%%%%%%%%%%%%%%%%%%%%%%%%%%%%%%%%%%%%%%%%%%%%%%%%%%%%%%%%%%%%%%%%%%%%%%%%%%%%
\section{FAST: Functional Adaptive Sensing and Testing}
\label{sec:method}

This section introduces FAST for QCD of functional data streams under limited sensing. Section~\ref{subsec:functional_modeling} presents a basis expansion model for representing sparse shifts in the latent coefficient space. Section~\ref{subsec:sparse_learning} develops a sliding-window Lasso estimator for online sparse learning. Section~\ref{subsec:adaptive_sensing} introduces a UCB-based bandit policy for adaptive sensing. Finally, Section~\ref{subsec:sequential_testing} presents the recursive likelihood-ratio testing procedure and the corresponding stopping rule.
To maintain consistency with the general formulation in Section~\ref{sec:problem_formulation}, we use the same detection-policy notation when presenting FAST in the remainder of the paper.

\subsection{Functional Data Modeling}
\label{subsec:functional_modeling}
We represent the functional data through a finite-dimensional basis expansion, a standard modeling device in functional data analysis that converts infinite-dimensional observations into latent coefficient dynamics \citep{fda,lei2023new}. Specifically, after centering the pre-change mean function, each observation is modeled as
\begin{equation}
\label{eq:basis_ex}
y_t(s)=\bphi(s)^\top\vx_t+e_t(s)
=\sum_{k=1}^{d}\phi_k(s)x_{tk}+e_t(s),
\quad s\in\mathcal S,
\end{equation}
where $\bphi(s)=[\phi_1(s),\ldots,\phi_d(s)]^\top$ is a collection of localized basis functions, $\vx_t=[x_{t1},\ldots,x_{td}]^\top$ is the latent coefficient vector, and $e_t(s)\sim\mathcal N(0,\sigma_e^2)$ is measurement noise. 

Following many established works in sequential change detection 
\citep{xie2013sequential,zhang2019partially}, we assume the coefficient vector $\vx_t$ follows a multivariate Gaussian distribution with mean vector $\btheta$ and covariance matrix $\bSigma$.
Thus, $\btheta$ carries the underlying
structural information of the functional data and serves as the target of online change detection. The Gaussian coefficient model yields closed-form likelihood ratios and KL information, which facilitates both the sequential testing and adaptive sensing design. For more general known distributions, the same framework can be applied by replacing these quantities accordingly. Additionally, the structural parameters ($\bphi(s)$, $\bSigma$, $\sigma_e^2$) are assumed to be known or can be reliably estimated from historical offline data, as detailed in Appendix A.

Due to practical sensing constraints, we cannot observe the full functional data $y_t(s)$ across $s\in\mathcal{S}$. Instead, at each time step $t$, we select a subset of sensing locations $Z_t \subset \mathcal{S}$ subject to a sensing budget $|Z_t| = m$. The subsampled observation vector $\vY_{Z_t} \in \mathbb{R}^m$ can be succinctly written in matrix form as:
\begin{equation*}
\label{eq:partial_fpca}
    \vY_{Z_t} = \bPhi_{Z_t} \vx_t + \ve_{Z_t},
\end{equation*}
where $\bPhi_{Z_t} \in \mathbb{R}^{m \times d}$ and $\ve_{Z_t} \sim \mathcal{N}(\mathbf{0}, \sigma_e^2 \vI)$ are the basis matrix and the noise vector restricted to the selected subset $Z_t$, respectively.
Building upon this limited-sensing setting, we formulate QCD as a sequential testing problem in which the coefficient mean $\btheta$ shifts at an unknown time $\tau_0$. The pre-change and post-change hypotheses are:
\begin{equation*}
    H_0: \btheta = \mathbf{0} \quad \text{for } t < \tau_0, \quad H_1: \btheta = \btheta_1 \neq \mathbf{0} \quad \text{for } t \ge \tau_0.
\end{equation*}

As established in Section \ref{sec:problem_formulation}, both the oracle sensing subset and the corresponding detection-delay benchmark depend on the post-change parameter $\btheta_1$. In practice, however, $\btheta_1$ is unknown and must be dynamically inferred from subsampled observations, creating a coupled learning-and-sensing problem. Consequently, to achieve rapid and reliable testing, we need to answer two key questions: (1) how to formulate a robust online estimation procedure for the sparse vector $\btheta_1$; and (2) how to develop an adaptive sensing policy to select the $m$ sensing locations, thereby maximizing detection power.

\begin{remark}
    Crucially, we assume that the post-change mean vector $\btheta_1$ is strictly sparse with $a = \|\btheta_1\|_0 \ll d$. This assumption aligns well with practice: real-world functional shifts are typically localized to small regions rather than spreading across the entire domain \citep{zhang2021dynamic,guo2023bayesian}. By using localized basis functions, a physical change naturally translates into a sparse change in the basis coefficients. However, sparse learning becomes more
challenging under adaptive sensing, because the sensing locations depend on past
observations, making the resulting data policy-dependent and generally
non-i.i.d. Consequently, established sparse learning results cannot be applied
directly, and new theoretical analysis is required.
\end{remark}

\subsection{Sparse Learning}
\label{subsec:sparse_learning}
In this section, we detail how to estimate $\btheta_1$ from subsampled observations. Instead of using all historical data, we estimate the parameter over a sliding window $\mathcal{W}_t=\{\max(1, t-w+1), \ldots, t\}$, which incorporates the $w$ most recent observations up to time $t$. The purpose is to mitigate contamination from stale data and rapidly track emerging sparse mean shifts. Under the basis expansion model in Equation (\ref{eq:basis_ex}), the subsampled observation at time $t$ satisfies $\vY_{Z_t} \sim \mathcal{N}\!\left(\bPhi_{Z_t}\btheta, \bOmega_{Z_t}\right)$, where $\bOmega_{Z_t} = \bPhi_{Z_t}\bSigma\bPhi_{Z_t}^{\top} + \sigma_e^2 \vI\in\mathbb R^{m\times m}$. Exploiting this Gaussian likelihood, we compute a sparse estimate $\hat{\btheta}_t$ by solving the Lasso problem, which corresponds to an $L_1$-penalized maximum likelihood estimation \citep{wang2018high,li2025feature}:
\begin{equation}
\label{eq:lasso_obj}
\begin{aligned}
\hat{\btheta}_t
&\in \arg\min_{\btheta\in\mathbb{R}^d}  \frac{1}{2w}\sum_{i\in\mathcal{W}_t} \left\| \bOmega_{Z_i}^{-1/2} \bigl(\vY_{Z_i}-\bPhi_{Z_i}\btheta\bigr) \right\|^2 + \lambda_t\|\btheta\|_1 \\
&=\arg\min_{\btheta\in\mathbb{R}^d} \frac{1}{2w}\btheta^\top \left( \sum_{i \in \mathcal{W}_t} \bPhi_{Z_i}^\top \bOmega_{Z_i}^{-1} \bPhi_{Z_i} \right) \btheta - \frac{1}{w} \left( \sum_{i \in \mathcal{W}_t} \bPhi_{Z_i}^\top \bOmega_{Z_i}^{-1} \vY_{Z_i} \right)^\top \btheta +  \lambda_t \|\btheta\|_1,
\end{aligned}
\end{equation}
where $\lambda_t$ is the regularization parameter that controls the sparsity of the estimate.

Let $\vS_Z = \bPhi_Z^\top \bOmega_Z^{-1} \bPhi_Z\in\mathbb R^{d\times d}$ denote the instantaneous information matrix for any candidate subset $Z$. We define the sufficient statistics $\vS_t := \sum_{i\in\mathcal{W}_t}\vS_{Z_i}$ and $\vb_t := \sum_{i\in\mathcal{W}_t}\bPhi_{Z_i}^{\top}\bOmega_{Z_i}^{-1}\vY_{Z_i}$, which are fundamental for solving Equation (\ref{eq:lasso_obj}) and admit computationally efficient recursions:
\begin{equation}
\label{eq:SB}
    \begin{aligned}
\vS_t &= \vS_{t-1} + \vS_{Z_t} - \vS_{Z_{t-w}}, \\
\vb_t &= \vb_{t-1} + \bPhi_{Z_t}^{\top}\bOmega_{Z_t}^{-1}\vY_{Z_t} - \bPhi_{Z_{t-w}}^{\top}\bOmega_{Z_{t-w}}^{-1}\vY_{Z_{t-w}},
\end{aligned}
\end{equation}
where we set $\vS_{Z_i}=\mathbf 0$ and $
\bPhi_{Z_i}^{\top}\bOmega_{Z_i}^{-1}\vY_{Z_i}=\mathbf 0$ for $i\le 0$. Using these recursively updated sufficient statistics, we compute $\hat{\btheta}_t$ via coordinate descent. At each time $t$, starting from the previous estimate or a warm-start initialization, the algorithm cycles through coordinates $k=1,\ldots,d$ until convergence. For each coordinate, one update takes the soft-thresholding form:
\begin{equation}
    \label{eq:theta_cd}
\hat{\btheta}_t(k) \leftarrow \frac{1}{\vS_t(k,k)} \, \mathrm{Soft} \!\left( \vb_t(k)-\sum_{j\neq k}\vS_t(k,j)\hat{\btheta}_t(j), \, w\lambda_t \right),
\end{equation}
where $\mathrm{Soft}(u,\lambda) = \mathrm{sign}(u)(|u|-\lambda)_+$. To quantify estimation uncertainty for adaptive sensing, we use the inverse $\vV_t=\vS_t^{-1}$ once the local information matrix is nonsingular; rank-deficient epochs use an initialization action (Appendix~B.4). The inverse is used to construct the ellipsoidal confidence set in
Section~\ref{subsec:adaptive_sensing}.

\subsection{Adaptive Sensing}
\label{subsec:adaptive_sensing}
Using the online estimate, we next determine the sensing subset $Z_{t+1}$ for the upcoming observation. This decision is naturally formulated within a CMAB framework \citep{chen2013combinatorial}. In this setting, each design point on the functional domain acts as an individual arm, and the learner must select a subset of arms subject to the sensing budget $|Z| =m$. Motivated by the minimax lower bound in Theorem~\ref{thm:lower_bound}, an effective sensing policy should accumulate post-change evidence at a rate as close as possible to the oracle KL information. Accordingly, the instantaneous KL information $\mathcal I(Z,\btheta_1)$ serves as the reward in our bandit formulation, leading to the following sensing objective:
\begin{equation}
    \label{eq:ideal_reward}
    Z_{t+1}^* = \arg\max_{Z: |Z|=m} \mathcal I(Z,\btheta_1) =\arg\max_{Z: |Z|=m} \frac{1}{2} \btheta_1^\top \vS_Z \btheta_1.
\end{equation}
%Evaluating Equation (\ref{eq:ideal_reward}) explicitly requires the true post-change parameter $\btheta_1$. However, because

Because $\btheta_1$ is unknown and must be estimated sequentially
from observations collected under bandit feedback, the learner faces
an exploration--exploitation tradeoff. Directly replacing $\btheta_1$
in Equation~(\ref{eq:ideal_reward}) with the current estimate
$\hat{\btheta}_t$ gives a purely exploitative rule, which may repeatedly
sample currently favored regions while overlooking potentially
informative locations. We therefore account for estimation uncertainty
through a UCB policy
\citep{abbasi2011improved,lattimore2020bandit}.
Specifically, we construct an ellipsoidal confidence set centered at the online Lasso estimate: 
\begin{equation} \label{eq:confidence} \mathcal{C}_t = \left\{ \btheta\in\mathbb{R}^d: \|\btheta-\hat{\btheta}_t\|_{\vV_t^{-1}} \leq\alpha_t \right\}, 
\end{equation} 
where $\alpha_t$ is a dynamically adjusted confidence radius, and $\|\cdot\|_{\vV_t^{-1}}$ is the weighted norm defined as $\|\vu\|_{\vV_t^{-1}}=(\vu^\top\vV_t^{-1}\vu)^{1/2}$.

To account for estimation uncertainty, we follow the principle of
optimism in the face of uncertainty and consider the following
sensing objective:
\begin{equation*}
\label{eq:optimistic_objective}
\max_{Z:\,|Z|=m}
\sup_{\btheta\in\mathcal C_t}
\mathcal I(Z,\btheta)
=
\max_{Z:\,|Z|=m}
\sup_{\btheta\in\mathcal C_t}
\frac{1}{2}\btheta^\top\vS_Z\btheta.
\end{equation*}

Rather than solving this nested optimization problem directly,
we derive a closed-form upper bound on its objective, leading
to the following tractable UCB-based sensing rule:
\begin{equation}
\label{eq:ucb_objective}
Z_{t+1}
\in
\arg\max_{Z:\,|Z|=m}
\overline{\mathcal I}_t(Z),
\end{equation}
where
\begin{equation*}
\begin{aligned}
\overline{\mathcal I}_t(Z)
:={}&
\underbrace{
\frac{1}{2}
\hat{\btheta}_t^\top
\vS_Z
\hat{\btheta}_t
}_{\text{Exploitation}}
+
\underbrace{
\alpha_t
\sqrt{
\hat{\btheta}_t^\top
\vS_Z
\vV_t
\vS_Z
\hat{\btheta}_t
}
+
\frac{\alpha_t^2}{2}
\left\|
\vV_t^{1/2}
\vS_Z
\vV_t^{1/2}
\right\|_{\mathrm{op}}
}_{\text{Exploration}}.
\end{aligned}
\end{equation*}
Here, $\|\cdot\|_{\mathrm{op}}$ denotes the matrix operator norm.
For every feasible sensing subset $Z$, on the event
$\btheta_1\in\mathcal C_t$, we have
$\mathcal I(Z,\btheta_1)
\leq
\sup_{\btheta\in\mathcal C_t}
\frac{1}{2}\btheta^\top\vS_Z\btheta
\leq
\overline{\mathcal I}_t(Z)$.
The first inequality follows from the inclusion
$\btheta_1\in\mathcal C_t$, while the second follows from
the Cauchy--Schwarz inequality and the operator norm bound.
Thus, $\overline{\mathcal I}_t(Z)$ provides a tractable
optimistic upper bound on the KL information, balancing
exploitation of the current estimate with exploration
under estimation uncertainty.

Obtaining the exact solution to Equation (\ref{eq:ucb_objective}) requires an exhaustive search over all possible combinations. To reduce complexity, we also introduce a greedy search alternative. Specifically, we iteratively select one point at a time to greedily maximize the UCB index \citep{joshi2008sensor}. Since the information matrix $\vS_Z$ is not necessarily diagonal, the arms cannot be evaluated independently; instead, we must systematically evaluate the marginal increase in the UCB for each arm. In particular, given a current subset $Z$ with $|Z|<m$, the next location is chosen as
\begin{equation}
\label{eq:greedy}
    s^* \in \arg\max_{s \notin Z}
    \left(\overline{\mathcal I}_t(Z\cup\{s\})-\overline{\mathcal I}_t(Z)\right).
\end{equation}

We then update $Z\leftarrow Z\cup\{s^*\}$ until $|Z|=m$. Under regularity conditions, this greedy rule admits a near-optimal approximation guarantee, as stated in Lemma~\ref{assump:greedy_error} \citep{elenberg2018restricted}.

\begin{lemma}[Greedy Approximation]
\label{assump:greedy_error}
Suppose that $\overline{\mathcal I}_t$ is normalized and
monotone, and its submodularity ratio is uniformly bounded
below by a constant $\zeta>0$.
Then the subset $Z_{t+1}$ selected by the greedy rule satisfies
\begin{equation}
\overline{\mathcal I}_t(Z_{t+1})
\geq
c\max_{Z:\,|Z|=m}\overline{\mathcal I}_t(Z),
\end{equation}
where $c=1-e^{-\zeta}$.
Under exact subset optimization, the same guarantee holds
with $c=1$.
\end{lemma}

\begin{remark}
Suppose the functional domain is discretized into $M$ candidate design points, denoted by $s_1,\ldots,s_M$. Exact optimization then requires evaluating all $\binom{M}{m}$ sensing subsets. In contrast, the greedy rank-one update selects sensing locations sequentially and requires only $\mathcal{O}(mM)$ objective evaluations. This reduces the search effort from combinatorial to linear growth and makes the sensing step suitable for online implementation.
\end{remark}

\subsection{Sequential Testing}
\label{subsec:sequential_testing}
Given the selected sensing subset $Z_{t+1}$, we observe $\vY_{Z_{t+1}}$ and update the following recursive likelihood-ratio statistic for sequential testing \citep{page1954continuous}:
\begin{equation}
\label{eq:test}
    W_{t+1} = \max(0, W_{t} + \ell_{t+1}),
\end{equation}
where $\ell_{t+1}$ is the plug-in log-likelihood ratio (LLR). To preserve
the non-anticipating structure, both $Z_{t+1}$ and
$\hat{\btheta}_{t}$ are determined from $\mathcal F_{t}$ before
$\vY_{Z_{t+1}}$ is observed. Specifically,
\begin{equation*}
    \ell_{t+1} = \log \frac{f(\vY_{Z_{t+1}} \mid \hat{\btheta}_t)}{f(\vY_{Z_{t+1}} \mid \mathbf{0})} = \hat{\btheta}_t^\top \bPhi_{Z_{t+1}}^\top \bOmega_{Z_{t+1}}^{-1} \vY_{Z_{t+1}} - \frac{1}{2} \hat{\btheta}_t^\top \vS_{Z_{t+1}} \hat{\btheta}_t.
\end{equation*}

FAST declares a change when the detection statistic exceeds a threshold $b$, yielding the stopping time $\tau_1=\inf\{t\ge 1: W_t>b\}$. In implementation, the threshold $b$ can be calibrated by Monte Carlo simulation to attain a target ARL$_0$ level \citep{keskin2021dynamic}. Theoretically, Lemma~\ref{lemma:threshold} shows that setting $b=\log\gamma$ guarantees admissibility. The detailed proof is deferred to Appendix B.3.

\begin{lemma}[Threshold Choice]
\label{lemma:threshold}
For any sensing policy $Z = \{Z_t\}_{t \ge 1}$ and target ARL$_0$ constraint $\gamma > 1$, setting the threshold $b = \log \gamma$ guarantees that the sequential test based on $W_t$ is admissible, i.e., $(Z, \tau_1) \in \mathcal{A}_{\gamma}$, which implies $\mathbb{E}_{\infty}[\tau_1] \ge \gamma$.
\end{lemma}

\begin{remark}
\label{remark1}
    When the system operates in the pre-change regime ($H_0$), the expected value of $\vY_{Z_{t+1}}$ is zero, inducing a negative drift in the expected LLR and keeping the statistic near zero. Conversely, under the post-change regime ($H_1$), as the parameter estimate becomes accurate, the expected LLR approaches $\frac{1}{2} \btheta_1^\top \vS_{Z_{t+1}} \btheta_1$, which is exactly the reward function used in adaptive sensing. Hence, sensing subsets with larger KL information allow the test to accumulate evidence more rapidly and reduce the detection delay.
\end{remark}

Together, online sparse learning, UCB-based adaptive sensing, and
recursive sequential testing constitute the FAST procedure for online decision making. Algorithm \ref{alg:fast} summarizes the complete FAST workflow. 

\begin{algorithm}[!h]
\caption{FAST: functional adaptive sensing and testing.}
\label{alg:fast}
\begin{algorithmic}[1]

\State Initialize $\hat{\btheta}_0=\mathbf 0$, $W_0=0$, and choose $Z_1$ randomly with $|Z_1|=m$

\For{$t=1,2,\ldots$}
\State Observe the subsampled data $\vY_{Z_t}$

\State Compute the detection statistic $W_t$ using Equation~(\ref{eq:test})

\If{$W_t>b$}
    \State Stop and declare a change; set $\tau_1=t$
\EndIf

\State Update the sliding-window sufficient statistics $\vS_t$ and $\vb_t$ via Equation (\ref{eq:SB})

\State Update the Lasso estimate $\hat{\btheta}_t$ using Equation~(\ref{eq:theta_cd}) (zero-information coordinates are set to zero)

\State Construct $\mathcal C_t$ using Equation~(\ref{eq:confidence}) when defined
\State Select $Z_{t+1}$ by Equation~(\ref{eq:ucb_objective}) or~(\ref{eq:greedy}) (initialization: Appendix~B.4)

\EndFor
\end{algorithmic}
\end{algorithm}

%%%%%%%%%%%%%%%%%%%%%%%%%%%%%%%%%%%%%%%%%%%%
\section{Theoretical Properties}
\label{sec:theory}
In this section, we rigorously establish that FAST achieves asymptotic optimality for QCD in
functional data streams. We first analyze the non-asymptotic error bound of the online Lasso estimator under limited sensing. Building upon this, we subsequently derive the asymptotic upper bound on the WADD. All detailed proofs are provided in Appendix B.

\subsection{Estimation Error Analysis}

We first quantify the estimation accuracy of the
online estimate $\hat{\btheta}_t$. Before presenting
the theoretical guarantee, we impose two standard
regularity conditions on the normalized accumulated
information matrix $w^{-1}\vS_t$ to ensure
sufficient curvature for sparse recovery.
Appendix~B.4 provides their formal definitions
and discusses their feasibility under partial sensing.

\begin{assumption}[Compatibility Condition]
\label{assum:compatibility}
The matrix $w^{-1}\vS_t$ satisfies the compatibility condition \citep{buhlmann2011statistics} with a constant $\kappa>0$.
\end{assumption}

\begin{assumption}[Restricted Eigenvalue Condition]
\label{assum:restricted_eigenvalue}
The matrix $w^{-1}\vS_t$ satisfies the restricted eigenvalue condition \citep{bickel2009simultaneous} with a constant $\rho>0$.
\end{assumption}

To bound the estimation error, the main technical difficulty is that, under adaptive sensing, the observations entering the Lasso problem are selected by the policy itself. Because the action $Z_t$ depends on the historical data $\mathcal F_{t-1}$ and only selected locations are observed under bandit feedback, the resulting observations are policy-dependent and no longer i.i.d. This prevents a direct application of classical oracle inequalities from high-dimensional regression and calls for a martingale-based analysis \citep{bastani2020online}.
In particular, Theorem~\ref{thm:consistency} establishes a non-asymptotic Euclidean error bound for the Lasso estimator. This result characterizes the effective sample complexity in Corollary~\ref{cor:budget}, which is essential in high-dimensional settings. It also supports the confidence ellipsoid in Corollary~\ref{cor:confidence}, which serves as a bridge from sparse learning to adaptive sensing. 

\begin{theorem}[Non-Asymptotic Error Bound]
\label{thm:consistency}
For any $t\ge \tau_0+w-1$, suppose the regularization parameter is set to $\lambda_t = 2\sigma_{\rm sc} \sqrt{\frac{2\log(2d/\delta)}{w}}$ where $\sigma_{\rm sc}^2=\max_{Z,k}\vS_Z(k,k)$. Then, for any $\delta \in (0, 1)$, with probability at least $1-\delta$, the Lasso estimator $\hat\btheta_t$ in Equation~(\ref{eq:theta_cd}) satisfies:
\begin{equation}
\|\hat\btheta_t-\btheta_1\|_2
\le
\frac{\sigma_{\rm sc}}{\kappa\sqrt{\rho w}}
\sqrt{72a\log(2d/\delta)}.
\end{equation}
\end{theorem}

\begin{remark}
The proof separates the estimation error into a deterministic sparse-recovery component and a random score-noise component. The latter is conditionally centered given the history and therefore forms a martingale difference sequence. A coordinate-wise conditional sub-Gaussian martingale bound, followed by a union bound, controls this term with high probability.
\end{remark}

\begin{corollary}[Effective Sample Complexity]
\label{cor:budget}
The sparse Lasso regularization reduces the effective learning dimension from $d$ to the sparse complexity $a\log d$. In particular, Theorem~\ref{thm:consistency} implies that $\hat\btheta_t$ is consistent in Euclidean norm when $a\log d=o(w)$.
\end{corollary}

\begin{corollary}[Confidence Ellipsoid]
\label{cor:confidence}
When $\vS_t$ is nonsingular, the proof of Theorem~\ref{thm:consistency} also yields the information-norm bound $\|\hat\btheta_t-\btheta_1\|_{\vV_t^{-1}}
\le
\frac{\sigma_{\rm sc}}{\kappa}\sqrt{72a\log(2d/\delta)}$
with probability at least $1-\delta$. Therefore, by setting $\alpha_t=
\frac{\sigma_{\rm sc}}{\kappa}\sqrt{72a\log(2d/\delta)}$, 
the ellipsoidal confidence set $\mathcal C_t$ constructed in Equation~(\ref{eq:confidence}) contains the true post-change parameter $\btheta_1$ with probability at least $1-\delta$.
\end{corollary}

%%%%%%%%%%%%%%%%%%%%%%%%%%%%%%%%%%%%%
\subsection{Detection Delay Analysis}
With the estimation guarantees established, we proceed to bound the WADD.
To translate parameter uncertainty into a bound on sensing performance,
we impose a mild local regularity condition on the KL information.
The first condition bounds the magnitude of the feasible post-change
signals, while the other ensures that the information from any single
sensing subset is controlled by the information accumulated over the
current sliding window. Under Assumption~\ref{assum:kl_regular} and the
high-probability coverage guarantee $\btheta_1\in\mathcal C_t$ in
Corollary~\ref{cor:confidence},
Lemma~\ref{prop:optimism_oracle} shows that the selected subset achieves
near-oracle KL information, up to the approximation factor $c$ and a vanishing
confidence error $\epsilon_{\mathrm{ucb}}$

%To link the sensing policy to the detection delay, we must analyze the KL information provided by the selected subset $Z_t$. Under above assumptions, Lemma \ref{prop:optimism_oracle} formalizes the optimism principle underlying UCB-type policies: whenever the true parameter lies within the confidence set, the selected subset enjoys near-oracle KL information, offset only by the greedy approximation factor and the confidence-set error $\epsilon_{\mathrm{ucb}}$ \citep{abbasi2011improved, lattimore2020bandit}. %Crucially, this optimism error is governed by the sliding window size $w$, shrinking smoothly as the window expands.

\begin{assumption}[Local KL Regularity]
\label{assum:kl_regular}
There exist constants $c_1,c_2>0$ such that, along the sensing
path generated by FAST and for all sufficiently large $t$,
\[
\|\btheta_1\|_{\vS_{Z_{t+1}}}\leq c_1,
\qquad
\left\|
\vV_t^{1/2}\vS_{Z_{t+1}}\vV_t^{1/2}
\right\|_{\mathrm{op}}
\leq\frac{c_2}{w}.
\]
\end{assumption}

\begin{lemma}[Near-Oracle KL Information]
\label{prop:optimism_oracle}
Conditioned on the high-probability event
$\btheta_1\in\mathcal C_t$, the instantaneous KL information of the
selected sensing subset satisfies
\[
\mathcal I(Z_{t+1},\btheta_1)
\geq
c\mathcal I^*(\btheta_1)-\epsilon_{\mathrm{ucb}},
\]
where, whenever $a\log(2d/\delta)=o(w)$,
$\epsilon_{\mathrm{ucb}}
=
\mathcal O\!\left(
\sqrt{\frac{a\log(2d/\delta)}{w}}
\right)$
is the optimism error induced by the confidence radius.
\end{lemma}

Based on Lemma \ref{prop:optimism_oracle}, we now present the main theorem bounding the WADD of FAST. 

\begin{theorem}[Detection Delay Bound]
\label{thm:upper_bound}
Set $b=\log\gamma$ and impose the uniform excitation, approximation, and moment conditions detailed in Appendix~B.7. As $w,\gamma\to\infty$ with $a\log(dw)=o(w)$ and $w=o(\log\gamma)$, the WADD of FAST satisfies:
\begin{equation}
    J_{\btheta_1}^{\rm FAST}(Z,\tau_1) \le \frac{\log \gamma}{c \mathcal{I}^*(\btheta_1)} (1 + o(1)).
\end{equation}
\end{theorem}

\begin{remark}
The proof splits the total delay into two phases: the initial window-filling period and the LLR accumulation phase. To bound the expected hitting time of the LLR process, we use a blockwise concentration argument for delayed-adapted increments (Appendix~B.7). As the window size $w \to \infty$, the optimism error and estimation error vanish, yielding an asymptotic lower bound of $c \mathcal{I}^*(\btheta_1)$ on the expected LLR drift. Consequently, the expected hitting time is bounded by the ratio of the threshold $\log \gamma$ to this asymptotic drift, while the initial window-filling time $w$ becomes negligible since $w = o(b)$.
\end{remark}

Comparing Theorem~\ref{thm:upper_bound} with the minimax lower bound in Theorem \ref{thm:lower_bound} yields the following direct corollary. In essence, FAST achieves an asymptotic detection
delay within a factor of $1/c$ of the minimax limit.

\begin{corollary}[First-Order Optimality] 
Under the conditions of Theorem~\ref{thm:upper_bound}, as $w,\gamma\to\infty$ with $a\log(dw)=o(w)$ and $w=o(\log\gamma)$, FAST satisfies
\begin{equation*}
\limsup_{\gamma\to\infty}
\frac{
J_{\btheta_1}^{\rm FAST}(Z,\tau_1)
}{
\inf_{(Z,\tau_1)\in\mathcal A_\gamma} J_{\btheta_1}(Z,\tau_1)
}
\le
\frac{1}{c}.
\end{equation*}

In particular, when the sensing subset is optimized exactly so that $c=1$, the ratio converges to one, and FAST is first-order asymptotically optimal.
\end{corollary}

\iffalse
\begin{corollary}[First-Order Optimality]
\label{cor:optimality}
If the inner maximization in Equation~(\ref{eq:ucb_objective}) is solved exactly, i.e., $c=1$, then the FAST policy is first-order asymptotically optimal:
\[
J_{\btheta_1}(Z,\tau_1)
\sim
\frac{\log\gamma}{\mathcal{I}^*(\btheta_1)}.
\]
More generally, under greedy subset selection with subset-optimization ratio $c\in(0,1]$, the policy is first-order optimal up to the factor $1/c$.
\end{corollary}
\fi

%%%%%%%%%%%%%%%%%%%%%%%%%%%%%%%%%%%%%%%%%%%%%%%%%%%%%%%%%%%%%%
\section{Simulations}
\label{sec:numerical}
This section evaluates FAST across four key dimensions: detection performance, online learning efficiency, parameter sensitivity, and empirical validation through a real-world case study.

\subsection{Experimental Setup}
\label{subsec:setup}

\textbf{\textit{Data Generation.}} 
We generate functional data streams $\{y_t(s):t=1,\ldots,1000\}$ on the one-dimensional interval $[0,1]$ using $d$ B-spline basis functions $\bphi(s)$ evaluated at $M$ equally spaced candidate points. The coefficient covariance matrix is set to $\bSigma(i,j)=(0.2)^{|i-j|}$, and the observation noise variance is fixed at $\sigma_e^2=0.01$. Under the pre-change regime, the mean coefficient vector is fixed at $\btheta_0=\mathbf{0}$, whereas after the change point $\tau_0=50$, we consider two sparse change patterns: Pattern I introduces a local functional mean shift with $y_t(s) \leftarrow y_t(s)+\Delta$ for $s\in\mathcal S_1$, and Pattern II introduces a coefficient shift with $\theta_{1,j_1}=\Delta$ and $\theta_{1,j_0}=0$ for all $j_0\neq j_1$. We consider three synthetic regimes with increasing problem sizes, denoted by Small, Medium, and Large; their detailed parameter configurations are summarized in Table~\ref{tab:exp_setting}.

\begin{table}[!h]
\centering
\caption{Experimental settings under different problem scales.}
\label{tab:exp_setting}
\scalebox{1}{
\begin{tabular}{lcccccc}
\hline
Scale & $M$ & $d$ & $m$ & $w$ & $\mathcal S_1$ for Pattern I & $j_1$ for Pattern II \\
\hline
Small  & 20  & 8  & 4  & 10 & $[0.10,0.35]$ & 3  \\
Medium & 100 & 20 & 10 & 15 & $[0.20,0.30]$ & 10 \\
Large  & 500 & 50 & 25 & 20 & $[0.40,0.44]$ & 25 \\
\hline
\end{tabular}}
\end{table}

\textbf{\textit{Benchmark Methods.}} 
We compare FAST with representative benchmarks covering optimism-based, posterior-based, and imputation-based adaptive sensing. Functional UCB (\textbf{FUCB}) is a simplified version of the proposed method that removes sparse regularization. Spatial rank-based augmentation sampling (\textbf{SRAS}) reconstructs missing entries and applies a rank-based detection statistic \citep{zan2023spatial}. \textbf{TS} adopts Thompson sampling with a smooth-sparse detection statistic \citep{guo2023bayesian}. \textbf{CMAB} uses UCB-based combinatorial sensing with an exponentially weighted moving average detection statistic \citep{zhang2019partially}. We also include an \textbf{Oracle} benchmark, which assumes known $\btheta_1$ and applies the proposed detector with oracle sensing.

\textbf{\textit{Evaluation Criteria and Implementation Details.}}
For each method, we calibrate the detection threshold using 2,000 pre-change replications so that the empirical $\mathrm{ARL}_0$ equals 200. We then compute the empirical average detection delay as $\mathrm{ADD}=R^{-1}\sum_{r=1}^{R}(\tau_1^{(r)}-\tau_0+1)^+$, where $R=1000$ is the number of post-change replications. To connect the empirical results with the theoretical analysis, we also report the online coefficient estimation error $\|\hat{\btheta}_t-\btheta_1\|_2$, the cumulative regret relative to the oracle subset $\sum_{t\ge\tau_0}[\mathcal I^*(\btheta_1)-\mathcal I(Z_t,\btheta_1)]$, and the runtime per time step.

\subsection{Detection Performance}
\label{subsec:mainresults}
\begin{figure*}[!h]
\centering
\subfloat[]{\includegraphics[width=0.32\linewidth]{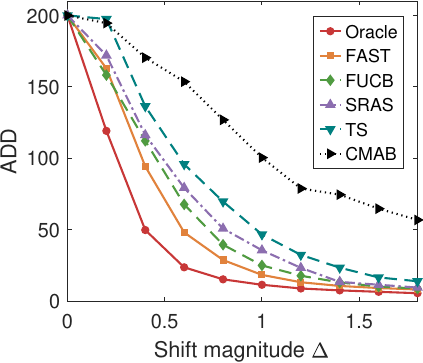}%
\label{fig:small_ab1}}
\hfil
\subfloat[]{\includegraphics[width=0.32\linewidth]{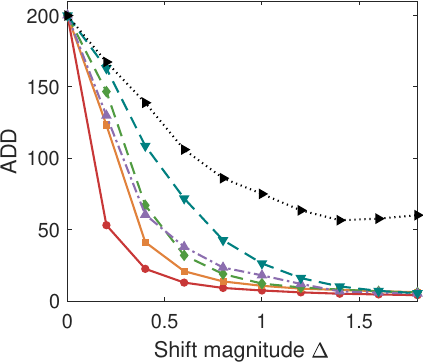}%
\label{fig:medium_ab1}}
\hfil
\subfloat[]{\includegraphics[width=0.32\linewidth]{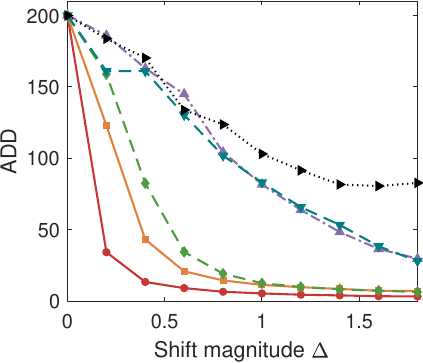}%
\label{fig:large_ab1}}

\subfloat[]{\includegraphics[width=0.32\linewidth]{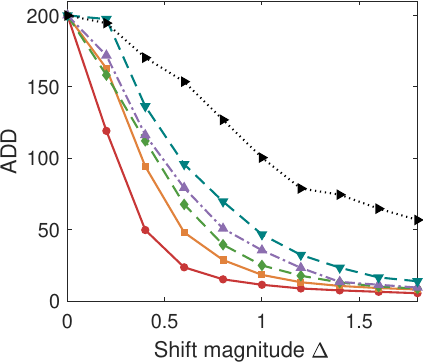}%
\label{fig:small_ab2}}
\hfil
\subfloat[]{\includegraphics[width=0.32\linewidth]{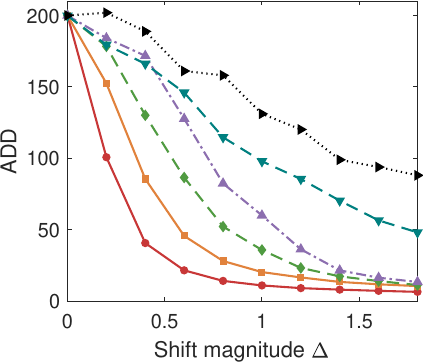}%
\label{fig:medium_ab2}}
\hfil
\subfloat[]{\includegraphics[width=0.32\linewidth]{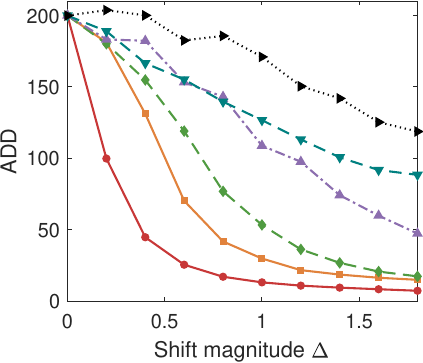}%
\label{fig:large_ab2}}
\caption{Detection performance comparison in terms of ADD under different scales and change patterns. (a) Small Scale, Pattern I. (b) Medium Scale, Pattern I. (c) Large Scale, Pattern I. (d) Small Scale, Pattern II. (e) Medium Scale, Pattern II. (f) Large Scale, Pattern II.}
\label{fig:scale_comparison}
\end{figure*}

Figure~\ref{fig:scale_comparison} illustrates the ADD of all competing methods across different shift magnitudes for the two change patterns under the three scale regimes. The standard deviations of detection delays are reported in Appendix C. As expected, the ADD of all methods generally decreases as the shift magnitude $\Delta$ increases. Across all scenarios, FAST delivers the most robust performance among implementable methods and closely tracks the Oracle benchmark. Its advantage becomes more significant in large-scale settings, where sparsity matters most. These results highlight the value of combining sparse online learning with UCB-based adaptive sensing. Without sparse regularization, FUCB becomes less effective in high-dimensional settings because it must learn the post-change direction over a larger parameter space. TS also shows slower detection, likely because posterior sampling allocates sensing resources more diffusely during the early post-change period. SRAS and CMAB perform worse overall, with performance further deteriorating as the dimension increases. This suggests that methods without explicit sparse coefficient modeling struggle to identify localized functional shifts under limited sensing.

\subsection{Online Learning Efficiency}
\begin{figure*}[!h]
\centering
\subfloat[]{\includegraphics[scale=0.68]{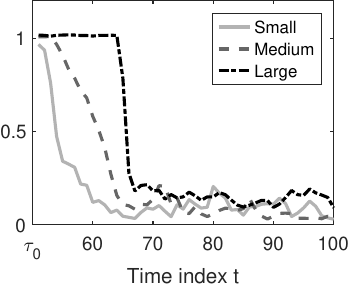}%
\label{fig:one_beta_err}}
\hfil
\subfloat[]{\includegraphics[scale=0.68]{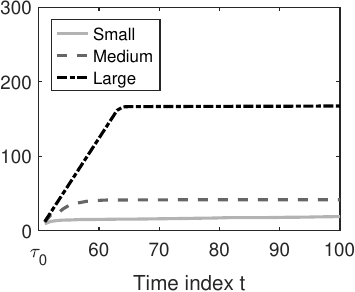}%
\label{fig:one_regret}}
\hfil
\subfloat[]{\includegraphics[scale=0.68]{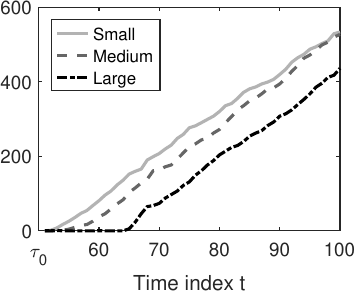}%
\label{fig:one_test_stat}}
\hfil
\subfloat[]{\includegraphics[scale=0.68]{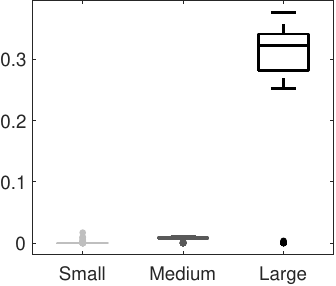}%
\label{fig:one_runtime}}
\caption{Performance metrics of FAST for a single simulation run. (a) Estimation Error. (b) Cumulative Regret. (c) Detection Statistic. (d) Runtime (seconds).}
\label{fig:one_rep_results}
\end{figure*}

To elucidate the mechanism behind the superior detection speed of FAST, we examine its online learning dynamics after the change occurs under Pattern II. Figure~\ref{fig:one_rep_results} summarizes the estimation quality, sensing efficiency, detection statistic, and runtime over the post-change horizon $\tau_0\le t\le 100$ in a representative simulation run. As shown in Figure~\ref{fig:one_beta_err}, the estimation error is initially large because the sliding window still contains pre-change observations. As more post-change samples enter the window, the estimator quickly adapts and the error declines, consistent with the online Lasso error bound in Theorem~\ref{thm:consistency}.

Correspondingly, Figure~\ref{fig:one_regret} shows that the regret growth gradually stabilizes as the sensing policy learns more informative regions. This aligns with the near-oracle KL information result in Lemma~\ref{prop:optimism_oracle}. Figure~\ref{fig:one_test_stat} further shows that, once the post-change parameter is estimated more accurately, the likelihood-ratio statistic accumulates evidence rapidly, consistent with the delay bound in Theorem~\ref{thm:upper_bound}. Finally, Figure~\ref{fig:one_runtime} shows that the runtime remains manageable even in the large-scale setting, because the recursive updates and greedy sensing rule avoid exhaustive combinatorial search.

\subsection{Parameter Sensitivity Analysis}
\label{subsec:sensitivity}
\begin{figure*}[!b]
\centering
\subfloat[]{\includegraphics[width=0.24\linewidth]{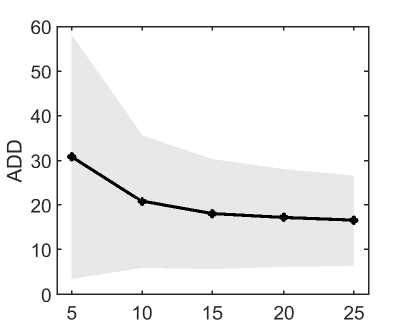}%
\label{fig:add_budget}}
\hfil
\subfloat[]{\includegraphics[width=0.24\linewidth]{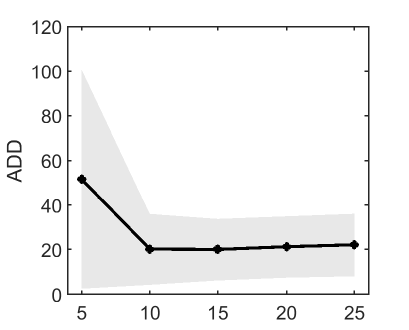}%
\label{fig:add_window}}
\hfil
\subfloat[]{\includegraphics[width=0.24\linewidth]{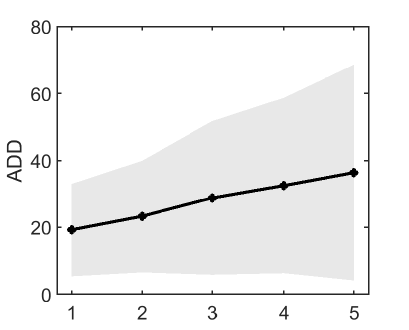}%
\label{fig:add_sparse}}
\hfil
\subfloat[]{\includegraphics[width=0.24\linewidth]{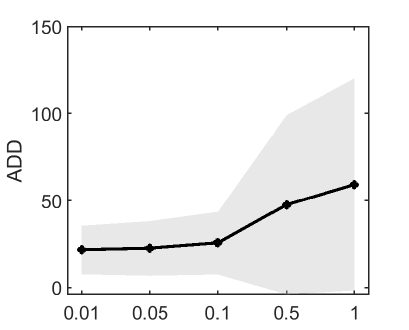}%
\label{fig:add_noise}}
\caption{Parameter sensitivity analysis of FAST in terms of ADD, with shaded bands denoting one standard deviation. (a) Budget $m$. (b) Window $w$. (c) Sparsity $a$. (d) Noise Variance $\sigma_e^2$.}
\label{fig:add_sensitivity}
\end{figure*}

We further examine the sensitivity of FAST to four key parameters, with one parameter varied at a time and the others fixed at the default configuration under the Medium-scale, Pattern II setting. Figure~\ref{fig:add_budget} shows that the ADD decreases as the sensing budget $m$ increases, since a larger budget provides more instantaneous information and increases the chance of sensing informative locations. However, the improvement becomes less pronounced when $m$ is sufficiently large, indicating diminishing returns after the most informative regions have been covered. Figure~\ref{fig:add_window} illustrates the effect of the sliding-window length $w$, which reflects a bias-variance tradeoff: a very short window leads to unstable sparse estimation, whereas an overly long window may retain stale pre-change observations and slow adaptation after the change. Figures~\ref{fig:add_sparse} and~\ref{fig:add_noise} show that the ADD increases with both the sparsity level $a$ and the noise variance $\sigma_e^2$. A larger $a$ corresponds to a less localized shift, making sparse recovery and informative sensing more difficult, which is consistent with the estimation error bound in Theorem~\ref{thm:consistency}; a larger $\sigma_e^2$ weakens the signal-to-noise ratio and reduces the effective KL information from each subsampled observation.

\subsection{Wildfire Detection Case Study}
\label{sec:case}

Wildfires have become a major threat to energy infrastructure, ecological systems, and public safety \citep{wei2025convolutional}. A notable example is the 2021 Dixie Fire in Northern California, which destroyed over 1,300 structures and caused an estimated USD $1.15$ billion in losses. Such incidents highlight the operational need for rapid wildfire detection under limited sensing resources. In practice, wildfire surveillance often relies on aerial inspection platforms, such as unmanned aerial vehicles (UAVs) and remote sensing systems, which can only observe a small portion of the spatial field at each decision epoch. This naturally leads to a functional change detection problem with limited sensing, where the fire-related signal evolves over a two-dimensional geographic domain, and the decision-maker must sequentially select only a limited number of locations to inspect.

\begin{figure}[!h]
  \centering
    \includegraphics[scale=0.65]{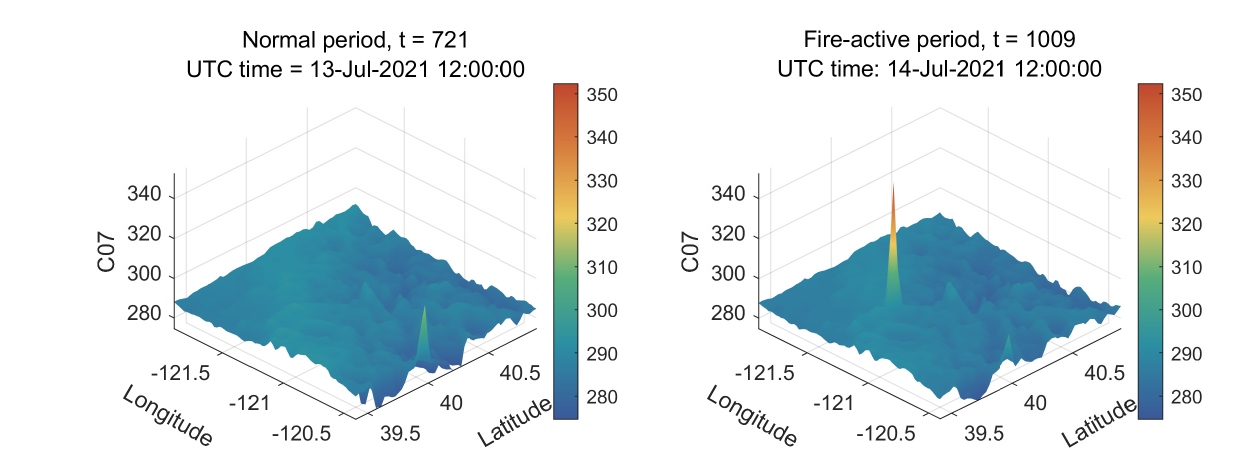}
    \caption{Representative GOES-17 C07 band observations.}
    \label{fig:case_data}
\end{figure}

We use GOES-17 satellite observations to validate the proposed method \citep{gerard2023wildfirespreadts}. The C07 band is selected because it is sensitive to thermal shifts associated with active fire behavior. We focus on a geographic region bounded by latitudes $39.4^\circ$N-$40.9^\circ$N and longitudes $121.9^\circ$W-$120.4^\circ$W, covering the early development area of the Dixie Fire. The observation period spans from July 11 to July 15. GOES-17 provides observations every five minutes with an approximately 2 km spatial resolution, yielding functional data $y_t(s)$ on a $56\times 73$ spatial grid for $t=1,\ldots,1440$. According to the event record, the ignition was associated with a tree falling and contacting power lines at approximately 13:48 UTC on July 13 ($t=743$). Around $\tau_0=764$, the accumulated heat produced a satellite-detectable thermal shift at grid location $s=(48,41)$, corresponding approximately to $39.96^\circ$N, $121.25^\circ$W. Figure~\ref{fig:case_data} shows representative pre-change and post-change functional observations.

For data preprocessing, we center each observation by subtracting the historical mean field for the same time period, thereby removing recurring baseline spatial patterns. To represent the functional data sequentially, we extract $d=100$ spatial basis functions from historical pre-change data via nonnegative matrix factorization (NMF). During online learning and detection, we use a window size of $w=20$ and impose a strict sensing budget of $m=10$ observation points per time step. We compare FAST against feasible adaptive benchmarks, with all methods calibrated to the same false-alarm constraint. SRAS and CMAB are excluded because they do not exploit sparsity or the latent functional structure represented in the coefficient space, leading to prohibitive computation time for this high-dimensional spatial field.

\begin{figure}[!h]
  \centering
    \includegraphics[scale=0.7]{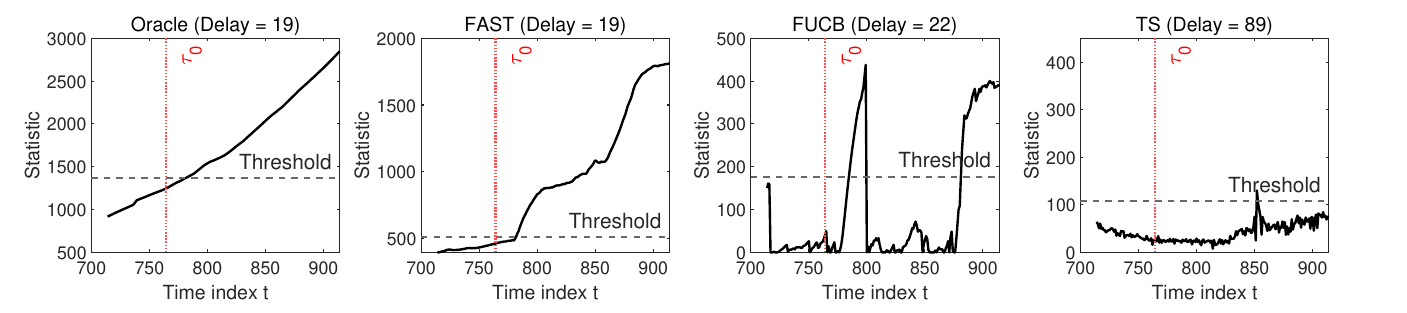}
    \caption{Detection statistic paths of the competing methods.}
    \label{fig:case_chart}
\end{figure}

Figure~\ref{fig:case_chart} reports the detection statistic paths. FAST raises the earliest alarm among the implementable methods at $\tau_1=782$, achieving a detection delay of $19$, which matches the Oracle benchmark. In contrast, FUCB exhibits a less stable statistic path because it does not explicitly exploit the sparsity of the post-change coefficient vector. TS performs worse because its posterior-sensing rule allocates the sensing budget more diffusely under noisy observations, slowing the accumulation of detection evidence.

\begin{figure}[!h]
  \centering
    \includegraphics[scale=0.5]{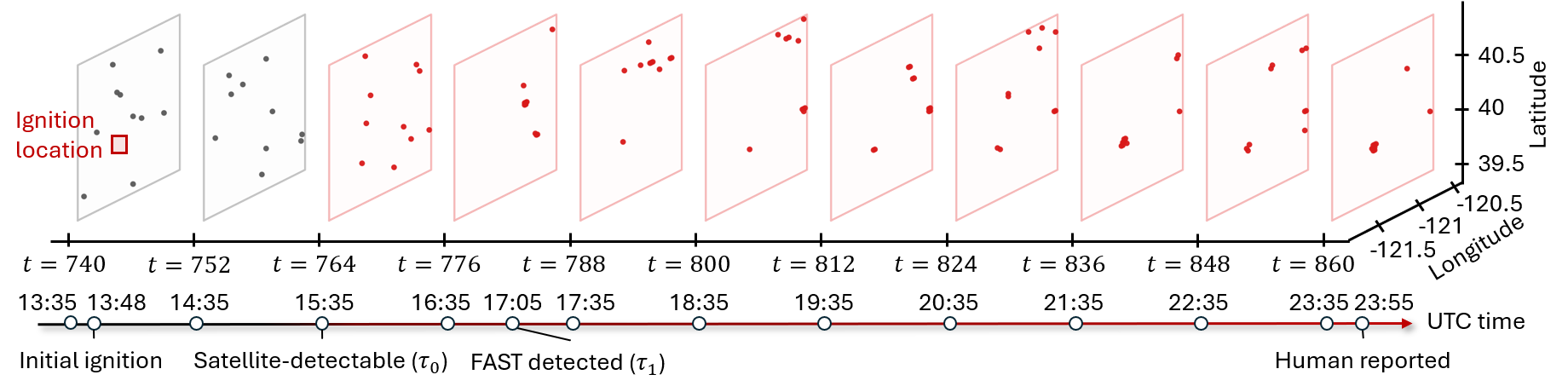}
    \caption{Selected sensing locations of FAST over time in the Dixie Fire case study.}
    \label{fig:case_sampling}
\end{figure}

Figure~\ref{fig:case_sampling} further illustrates the dynamic sensing behavior of FAST. Before the thermal shift becomes detectable, the selected sensing locations are widely dispersed across the region. Once the thermal shift emerges, the policy rapidly reallocates observations toward the post-change region near the ignition location. This transition from broad exploration to focused sensing shows that FAST adaptively concentrates sensing resources on informative spatial locations after the change occurs. Practically, FAST detects the abnormal thermal signal more than six hours before the first confirmed human observation of the fire, providing lead time for field verification, resource dispatch, and suppression planning.

%%%%%%%%%%%%%%%%%%%%%%%%%%%%%%%%%%%%%%%%%%%%%%%%%%%%%%%%%%%%%%%%%%%%%%%%%%%%%%%%%%%%%%%%%%%%%%
\section{Conclusion}
\label{sec:conclusion}
This paper studies the quickest change detection problem for partially observed functional data streams. At each time, only a limited number of locations on the functional domain can be observed, while the post-change parameter is unknown. We develop FAST, a Functional Adaptive Sensing and Testing procedure that combines a localized functional representation, sliding-window sparse estimation, adaptive sensing, and recursive likelihood-ratio testing. Localized functional changes are represented as sparse coefficient shifts, enabling the sensing policy to learn the post-change structure and progressively concentrate observations on informative locations.

FAST is supported by theoretical guarantees for estimation, sensing, and detection. We derive a minimax lower bound on the worst-case average detection delay, establish a non-asymptotic Lasso error bound under the policy-dependent observations induced by adaptive sensing, and show that the selected sensing subsets collect near-oracle KL information under the stated regularity conditions. The resulting sequential test achieves an asymptotic
detection delay within a factor of $1/c$ of the minimax
limit, and is first-order asymptotically optimal
under exact subset optimization. The threshold choice provides false-alarm control, while the greedy sensing rule yields a scalable implementation. Simulations and the GOES-17 wildfire case study illustrate the detection performance of FAST under limited sensing.
 
%This work also extends Lasso bandit learning beyond its conventional role in stationary regret minimization. In change detection, observations before and after the change follow different regimes, so aggregating all historical data is no longer appropriate. By incorporating a sliding-window Lasso estimator, FAST adapts sparsity-aware online learning to this nonstationary detection setting and uses recent observations to balance fast adaptation with stable estimation. By explicitly aligning the bandit reward with the drift of the LLR testing statistic, it further leverages a UCB policy to adaptively allocate sensing locations under estimation uncertainty. Theoretically, we derive a minimax lower bound on the worst-case detection delay, establish a non-asymptotic error bound for the online estimator, and show that the proposed policy attains a matching asymptotic upper bound. We also provide practical guarantees for online implementation, including false-alarm control, scalable greedy sensing, and regularity conditions for reliable sparse learning. Empirically, simulations and the GOES-17 wildfire case study demonstrate that FAST achieves shorter detection delays than representative benchmarks under the same false-alarm constraint.

Several extensions are worth pursuing. First, the Gaussian basis-coefficient model can be extended to non-Gaussian, heavy-tailed, or nonlinear functional data. Second, future work may consider more complex change patterns, including multiple changes, transient shifts, and time-varying post-change signals. Third, the known-sparsity assumption can be relaxed by developing sparsity-adaptive or tuning-free online detection policies.

\section*{Conflict of Interest}
The authors declare that they have no conflicts of interest.

\clearpage
\appendices
\renewcommand{\thesubsection}{\thesection.\arabic{subsection}}
\renewcommand{\theequation}{\thesection.\arabic{equation}}
\renewcommand{\thetable}{\thesection.\arabic{table}}
\renewcommand{\thefigure}{\thesection.\arabic{figure}}
\makeatletter
\gdef\thesubsectiondis{\thesection.\arabic{subsection}}
\gdef\thesubsubsectiondis{\thesubsectiondis.\arabic{subsubsection}}
\@addtoreset{equation}{section}
\@addtoreset{table}{section}
\@addtoreset{figure}{section}
\makeatother
\section{Offline Estimation of Structural Parameters}
\label{app:offline}

The online procedure treats the basis functions $\bphi(s)$, the coefficient covariance matrix $\bSigma$, and the measurement-noise variance $\sigma_e^2$ as structural quantities estimated from historical data. Let $\mathcal O_{\mathrm{off}}$ and $\mathcal O_{\mathrm{off}}^c$ denote, respectively, the sets of observed and missing entries in an incomplete historical data matrix $\vO\in\mathbb R^{T\times M}$, and let $\mathcal P_{\mathcal O_{\mathrm{off}}}$ and $\mathcal P_{\mathcal O_{\mathrm{off}}^c}$ be the corresponding projection operators. We estimate the latent functional representation through the smooth-sparse FPCA problem
\begin{equation}
\begin{aligned}
\min_{\vA,\bPhi,\vY}\quad
&\|\vY-\vY\bPhi\vA^\top\|_F^2
 +\lambda_1\|\bPhi\|_1
 +\lambda_2\|\vC\bPhi\|_F^2,\\
\text{s.t.}\quad
&\vA^\top\vA=\vI_d,
\qquad
\mathcal P_{\mathcal O_{\mathrm{off}}}(\vY)=\mathcal P_{\mathcal O_{\mathrm{off}}}(\vO),
\end{aligned}
\label{eq:offline_fpca}
\end{equation}
where $\bPhi=[\bphi_1,\ldots,\bphi_d]\in\mathbb R^{M\times d}$ contains the discretized basis functions evaluated at the $M$ candidate locations, $\vA\in\mathbb R^{M\times d}$ is an auxiliary orthonormal matrix, and $\vC$ is a prespecified roughness-penalty operator.

We solve Equation~\eqref{eq:offline_fpca} by block coordinate descent. Given the current iterates, the orthogonal factor is updated from the singular value decomposition $(\vY^{(r)})^\top\vY^{(r)}\bPhi^{(r)}=\widetilde{\vU}\widetilde{\bLambda}\widetilde{\vQ}^{\top}$ as $\vA^{(r+1)}=\widetilde{\vU}\widetilde{\vQ}^{\top}$. Conditional on the remaining components, each column $\bphi_k$ is updated by the convex problem $\min_{\bphi_k}\|\vr_k-\vD_k\bphi_k\|_2^2+\lambda_1\|\bphi_k\|_1+\lambda_2\|\vC\bphi_k\|_2^2$, where $\vr_k$ and $\vD_k$ are the corresponding partial residual and design matrix. The missing entries are then updated by $\vY^{(r+1)}=\mathcal P_{\mathcal O_{\mathrm{off}}}(\vO)+\mathcal P_{\mathcal O_{\mathrm{off}}^c}\{\vY^{(r)}\bPhi^{(r+1)}(\vA^{(r+1)})^\top\}$. After convergence, the estimated basis functions determine $\bPhi_{Z_t}$ in the online procedure, while $\bSigma$ and $\sigma_e^2$ are estimated from the corresponding coefficient scores and reconstruction residuals.

\section{Proofs}
\label{app:proofs}

\subsection{Proof of Theorem 1}
\begin{proof}
Fix a post-change parameter $\btheta_1$ such that $0<\mathcal I^*(\btheta_1)<\infty$. For a sensing policy $Z=\{Z_t\}_{t\ge1}$, define the true post-change log-likelihood-ratio increment
\begin{equation}
\ell_t^{(1)}
=
\log\frac{f(\vY_{Z_t}\mid\btheta_1)}{f(\vY_{Z_t}\mid\mathbf 0)}
=
\btheta_1^\top\bPhi_{Z_t}^\top\bOmega_{Z_t}^{-1}\vY_{Z_t}
-
\frac12\btheta_1^\top\vS_{Z_t}\btheta_1.
\label{eq:true_llr_increment}
\end{equation}
Because $Z_t$ is $\mathcal F_{t-1}$-measurable, under the post-change law,
\begin{equation}
\mathbb E_{\tau_0,\btheta_1}
\!\left[\ell_t^{(1)}\mid\mathcal F_{t-1}\right]
=
\mathcal I(Z_t,\btheta_1)
\le
\mathcal I^*(\btheta_1).
\label{eq:oracle_information_upper}
\end{equation}
Moreover, conditional Gaussianity gives
\begin{equation}
\operatorname{Var}_{\tau_0,\btheta_1}
\!\left(\ell_t^{(1)}\mid\mathcal F_{t-1}\right)
=
\btheta_1^\top\vS_{Z_t}\btheta_1
=
2\mathcal I(Z_t,\btheta_1)
\le
2\mathcal I^*(\btheta_1).
\label{eq:true_llr_variance}
\end{equation}
Thus, after centering the increments in Equation~\eqref{eq:true_llr_increment}, the resulting partial sums form a square-integrable martingale whose predictable quadratic variation grows at most linearly. Doob's inequality then implies that, for every $\varepsilon>0$,
\begin{equation}
\sup_{\tau_0\ge1}\operatorname*{ess\,sup}
\mathbb P_{\tau_0,\btheta_1}
\left(
\max_{1\le k\le n}
\sum_{i=\tau_0}^{\tau_0+k-1}\ell_i^{(1)}
\ge
(1+\varepsilon)n\mathcal I^*(\btheta_1)
\,\middle|\,
\mathcal F_{\tau_0-1}
\right)
\longrightarrow0.
\label{eq:llr_maximal_condition}
\end{equation}
Indeed, the deterministic sum of the conditional means is at most $n\mathcal I^*(\btheta_1)$ by Equation~\eqref{eq:oracle_information_upper}, whereas the probability that the centered martingale exceeds $\varepsilon n\mathcal I^*(\btheta_1)$ is of order $n^{-1}$ by Equation~\eqref{eq:true_llr_variance}.

Equation~\eqref{eq:llr_maximal_condition} is the standard maximal-information condition used in Lorden-type lower bounds; see, for example, \citet{veeravalli2024quickest}. The usual change-of-measure argument for any procedure satisfying $\mathbb E_\infty[\tau_1]\ge\gamma$ therefore yields, for every $\varepsilon>0$,
\begin{equation}
J_{\btheta_1}(Z,\tau_1)
\ge
\frac{(1-\varepsilon)\log\gamma}{\mathcal I^*(\btheta_1)}
\{1+o(1)\}.
\label{eq:lorden_lower_intermediate}
\end{equation}
Letting $\varepsilon\downarrow0$ and then taking the infimum over $(Z,\tau_1)\in\mathcal A_\gamma$ proves Theorem~1.
\end{proof}

\subsection{Proof of Lemma 1}
\begin{proof}
Fix a decision epoch $t$ and let $F_t(A)=\overline{\mathcal I}_t(A)$ for subsets of the finite candidate set $\mathcal M$, with $F_t(\varnothing)=0$. The submodularity-ratio hypothesis used here has the following precise meaning: there exists a constant $\zeta\in(0,1]$, independent of $t$, such that
\begin{equation}
\sum_{s\in B}\{F_t(A\cup\{s\})-F_t(A)\}
\ge \zeta\{F_t(A\cup B)-F_t(A)\}
\label{eq:weak_submodularity_ratio}
\end{equation}
for every pair of disjoint subsets $A,B\subseteq\mathcal M$ satisfying $|A|\le m-1$ and $1\le |B|\le m$. In particular, the hypothesis covers unions $A\cup B$ of cardinality up to $2m-1$; it is not restricted to sets whose union has cardinality at most $m$.

Write $G_k$ for the greedy set after $k$ additions, with $G_0=\varnothing$, and let $Z^\star\in\arg\max_{|Z|=m}F_t(Z)$. Put $B_k=Z^\star\setminus G_k$. By monotonicity and \eqref{eq:weak_submodularity_ratio},
\begin{align}
F_t(Z^\star)-F_t(G_k)
&\le F_t(G_k\cup Z^\star)-F_t(G_k)\notag\\
&\le \zeta^{-1}\sum_{s\in B_k}\{F_t(G_k\cup\{s\})-F_t(G_k)\}\notag\\
&\le \frac{m}{\zeta}\{F_t(G_{k+1})-F_t(G_k)\}.
\label{eq:greedy_one_step}
\end{align}
Consequently, with $D_k=F_t(Z^\star)-F_t(G_k)\ge0$,
$D_{k+1}\le (1-\zeta/m)D_k$. Since $F_t(G_0)=0$, iteration over $k=0,\ldots,m-1$ gives
\begin{equation}
F_t(G_m)\ge \{1-(1-\zeta/m)^m\}F_t(Z^\star)
\ge (1-e^{-\zeta})\max_{|Z|=m}F_t(Z).
\label{eq:greedy_approx_appendix}
\end{equation}
Exact maximization of $F_t$ gives the same inequality with $c=1$ and requires neither monotonicity nor weak submodularity.

For the three-term UCB index in Equation~\eqref{eq:ucb_objective}, neither monotonicity nor a uniform positive submodularity ratio follows from the Loewner ordering of the KL information matrices alone: in particular, the cross term $\sqrt{\hat{\btheta}_t^\top\vS_Z\vV_t\vS_Z\hat{\btheta}_t}$ need not be monotone. Thus the greedy factor $c=1-e^{-\zeta}$ holds under the stated assumptions, while exact UCB maximization satisfies the approximation inequality with $c=1$ without them.
\end{proof}

\subsection{Proof of Lemma 2}
\begin{proof}
Let $L_t=\exp(\ell_t)$. Because $Z_t$ and $\hat{\btheta}_{t-1}$ are $\mathcal F_{t-1}$-measurable,
\begin{equation}
\mathbb E_\infty[L_t\mid\mathcal F_{t-1}]
=
\int_{\mathbb R^m}
\frac{f(\vy\mid\hat{\btheta}_{t-1},Z_t)}{f(\vy\mid\mathbf0,Z_t)}
 f(\vy\mid\mathbf0,Z_t)\,d\vy
=1.
\label{eq:lr_mean_one}
\end{equation}
Define $R_0=0$ and
$R_t=(1+R_{t-1})L_t$. Then
\[
R_t=\sum_{k=1}^t\prod_{i=k}^tL_i,
\qquad
\mathbb E_\infty[R_t\mid\mathcal F_{t-1}]=R_{t-1}+1,
\]
so $R_t-t$ is a martingale under $\mathbb P_\infty$. Optional stopping at $\tau_1\wedge n$ gives
$\mathbb E_\infty[R_{\tau_1\wedge n}]=\mathbb E_\infty[\tau_1\wedge n]$.
If $\mathbb P_\infty(\tau_1=\infty)>0$, then $\mathbb E_\infty[\tau_1]=\infty$ and the result is immediate. Otherwise, Fatou's lemma and monotone convergence yield
\begin{equation}
\mathbb E_\infty[\tau_1]
\ge
\mathbb E_\infty[R_{\tau_1}].
\label{eq:sr_optional_stopping}
\end{equation}
The recursion $W_t=\max(0,W_{t-1}+\ell_t)$ implies
\[
e^{W_t}
=
\max\left\{1,\max_{1\le k\le t}\prod_{i=k}^tL_i\right\}
\le
1+R_t.
\]
At the stopping time, $W_{\tau_1}>b$, and one of the products in $R_{\tau_1}$ therefore exceeds $e^b$. Hence $R_{\tau_1}>e^b$. Combining this fact with Equation~\eqref{eq:sr_optional_stopping} gives
$\mathbb E_\infty[\tau_1]>e^b$. Setting $b=\log\gamma$ proves
$\mathbb E_\infty[\tau_1]\ge\gamma$ and therefore
$(Z,\tau_1)\in\mathcal A_\gamma$.
\end{proof}

\subsection{Regularity Conditions for Sparse Estimation}
\label{app:regularity}

Let $\mathcal B=\operatorname{supp}(\btheta_1)$, with $|\mathcal B|=a$, and denote its complement by $\mathcal B^c$. Define the cone
\begin{equation}
\mathcal K(\mathcal B)=\{\vu\in\mathbb R^d:\|\vu_{\mathcal B^c}\|_1\le3\|\vu_{\mathcal B}\|_1\},
\label{eq:restricted_cone}
\end{equation}
and let $\bar{\vS}_t=w^{-1}\vS_t$. The compatibility and restricted-eigenvalue conditions in Assumptions~1--2 require, for all $\vu\in\mathcal K(\mathcal B)$,
\begin{align}
\|\vu_{\mathcal B}\|_1^2
&\le \frac{a}{\kappa^2}\vu^\top\bar{\vS}_t\vu,
\label{eq:compatibility_appendix}\\
\vu^\top\bar{\vS}_t\vu
&\ge \rho\|\vu\|_2^2.
\label{eq:re_appendix}
\end{align}
These conditions control estimation error only on the sparse cone, so they do not require $\vS_t$ to be invertible. Accordingly, the proof of Theorem~2 uses the information seminorm $\|\vu\|_{\vS_t}=(\vu^\top\vS_t\vu)^{1/2}$, which is well-defined even for singular $\vS_t$.

The inverse-based UCB in Equation~\eqref{eq:ucb_objective}, however, is used only when $\vS_t\succ0$. This distinction is necessary: if a null-space direction of $\vS_t$ carries positive information for some candidate action, the supremum of its KL reward over the associated seminorm confidence set is unbounded. Substituting the Moore--Penrose pseudoinverse into the UCB formula would not yield a valid optimistic upper bound in that case. When $\vS_t(k,k)=0$, positive semidefiniteness also implies that the $k$th row and column of $\vS_t$ and the corresponding component $\vb_t(k)$ vanish; with $\lambda_t>0$, the Lasso minimizer sets $\hat{\btheta}_t(k)=0$ without performing the division in Equation~\eqref{eq:theta_cd}.

A predictable spanning design can be used whenever the accumulated matrix is rank deficient, provided the candidate basis evaluations span $\mathbb R^d$ and the window can accommodate a spanning collection of actions. Such a design preserves the likelihood-ratio martingale argument in Appendix~B.3. Since a sliding window may subsequently lose rank, the full-rank property and the uniform compatibility, restricted-eigenvalue, local-KL, and subset-approximation conditions needed for Theorem~3 are imposed along the mature post-change sensing path. They are not consequences of the initialization convention alone.

The sparse curvature conditions do not themselves entail full-space excitation. For instance, if $\bar{\vS}_t$ is block diagonal with respect to $\mathcal B$ and $\mathcal B^c$, with $\lambda_{\min}(\bar{\vS}_{t,\mathcal B\mathcal B})\ge\underline s_{\mathcal B}>0$ and a positive-semidefinite complementary block, then \eqref{eq:compatibility_appendix} holds with $\kappa^2=\underline s_{\mathcal B}$. Moreover, the cone inequality $\|\vu\|_2^2\le(1+9a)\|\vu_{\mathcal B}\|_2^2$ gives \eqref{eq:re_appendix} with $\rho=\underline s_{\mathcal B}/(1+9a)$, whether or not the complementary block is nonsingular. The separate full-rank requirement arises from the ellipsoidal UCB construction rather than from sparse recovery.

\subsection{Proof of Theorem 2}
\begin{proof}
Let $\vh=\hat{\btheta}_t-\btheta_1$. For $t\ge\tau_0+w-1$, every observation in $\mathcal W_t$ follows the post-change model. Define
\[
\bxi_i
=
\vY_{Z_i}-\bPhi_{Z_i}\btheta_1,
\qquad
\vg_t
=
\frac1w\sum_{i\in\mathcal W_t}
\bPhi_{Z_i}^\top\bOmega_{Z_i}^{-1}\bxi_i.
\]
Let $\ve_k$ denote the $k$th canonical basis vector in $\mathbb R^d$. Conditional on $\mathcal F_{i-1}$, $\bxi_i$ is centered Gaussian with covariance $\bOmega_{Z_i}$. Therefore, for every scalar $\nu\in\mathbb R$ and every $k=1,\ldots,d$,
\begin{align}
&\mathbb E_{\tau_0,\btheta_1}\!\left[
\exp\left\{
\nu\,\ve_k^\top
\bPhi_{Z_i}^\top\bOmega_{Z_i}^{-1}\bxi_i
\right\}
\,\middle|\,
\mathcal F_{i-1}
\right]
\notag\\
&\qquad =
\exp\left\{
\frac{\nu^2}{2}\vS_{Z_i}(k,k)
\right\}
\le
\exp\left\{
\frac{\nu^2\sigma_{\rm sc}^2}{2}
\right\},
\label{eq:score_coordinate_mgf}
\end{align}
where
\[
\sigma_{\rm sc}^2
:=
\max_{\substack{Z:\,|Z|=m\\1\le k\le d}}
\vS_Z(k,k).
\]
Iterating this conditional moment-generating-function bound over the window and applying a union bound over the $d$ coordinates gives
\begin{equation}
\mathbb P_{\tau_0,\btheta_1}\left(
\|\vg_t\|_\infty
>
\sigma_{\rm sc}
\sqrt{\frac{2\log(2d/\delta)}{w}}
\,\middle|\,
\mathcal F_{t-w}
\right)
\le
\delta.
\label{eq:score_concentration}
\end{equation}

Optimality of $\hat{\btheta}_t$ in Equation~(2) of the main text, together with
$\vY_{Z_i}-\bPhi_{Z_i}\hat{\btheta}_t
=\bxi_i-\bPhi_{Z_i}\vh$, yields
\begin{equation}
\frac{1}{2w}\|\vh\|_{\vS_t}^2
+
\lambda_t\|\hat{\btheta}_t\|_1
\le
\vh^\top\vg_t
+
\lambda_t\|\btheta_1\|_1.
\label{eq:basic_ineq}
\end{equation}
Set
$\lambda_t=2\sigma_{\rm sc}\sqrt{2\log(2d/\delta)/w}$ and work on the event
$\mathcal E_t=\{\|\vg_t\|_\infty\le\lambda_t/2\}$, which has conditional probability at least $1-\delta$. By Hölder's inequality and decomposability of the $\ell_1$ norm,
\begin{equation}
\frac{1}{2w}\|\vh\|_{\vS_t}^2
+
\frac{\lambda_t}{2}\|\vh_{\mathcal B^c}\|_1
\le
\frac{3\lambda_t}{2}\|\vh_{\mathcal B}\|_1.
\label{eq:cone_basic_inequality}
\end{equation}
Hence $\vh\in\mathcal K(\mathcal B)$. Applying Equation~\eqref{eq:compatibility_appendix} gives
\[
\|\vh_{\mathcal B}\|_1
\le
\frac{\sqrt a}{\kappa\sqrt w}
\|\vh\|_{\vS_t}.
\]
Substituting this bound into Equation~\eqref{eq:cone_basic_inequality} and dividing by
$\|\vh\|_{\vS_t}$ when it is nonzero yields
\begin{equation}
\|\hat{\btheta}_t-\btheta_1\|_{\vS_t}
\le
\frac{\sigma_{\rm sc}}{\kappa}
\sqrt{72a\log(2d/\delta)}.
\label{eq:info_norm_bound}
\end{equation}
This is the information-seminorm statement. When $\vS_t\succ0$, we have $\vV_t^{-1}=\vS_t$ and it is precisely the confidence-radius statement used in Corollary~2. Finally, Equation~\eqref{eq:re_appendix} implies
\[
\|\vh\|_{\vS_t}^2
=w\vh^\top\bar{\vS}_t\vh
\ge
\rho w\|\vh\|_2^2.
\]
Combining this relation with Equation~\eqref{eq:info_norm_bound} gives
\begin{equation}
\|\hat{\btheta}_t-\btheta_1\|_2
\le
\frac{\sigma_{\rm sc}}{\kappa\sqrt{\rho w}}
\sqrt{72a\log(2d/\delta)},
\label{eq:l2_error_bound_appendix}
\end{equation}
which proves Theorem~2.

The order statement in Corollary~1 follows immediately from Equation~\eqref{eq:l2_error_bound_appendix}: for any sequence $\delta_w\downarrow0$ satisfying
$a\log(2d/\delta_w)=o(w)$, the right-hand side converges to zero with probability at least $1-\delta_w$. When $\vS_t\succ0$, Corollary~2 follows from Equation~\eqref{eq:info_norm_bound} by taking
$\alpha_t=(\sigma_{\rm sc}/\kappa)\sqrt{72a\log(2d/\delta)}$.
\end{proof}

\subsection{Proof of Lemma 3}
\begin{proof}
The proof is carried out at an epoch $t$ for which $\vS_t\succ0$, so that $\vV_t=\vS_t^{-1}$ and the confidence set in Equation~\eqref{eq:confidence} is a bounded ellipsoid. Take an arbitrary candidate $Z$ and write $\btheta=\hat{\btheta}_t+\vu$ with $\|\vu\|_{\vS_t}\le\alpha_t$. Expanding the quadratic KL information and applying Cauchy--Schwarz in the pair of inverse information norms gives
\begin{align}
\mathcal I(Z,\btheta)
&=\frac12\hat{\btheta}_t^\top\vS_Z\hat{\btheta}_t
+\vu^\top\vS_Z\hat{\btheta}_t
+\frac12\vu^\top\vS_Z\vu\notag\\
&\le\frac12\hat{\btheta}_t^\top\vS_Z\hat{\btheta}_t
+\alpha_t\sqrt{\hat{\btheta}_t^\top\vS_Z\vV_t\vS_Z\hat{\btheta}_t}
+\frac{\alpha_t^2}{2}
\|\vV_t^{1/2}\vS_Z\vV_t^{1/2}\|_{\rm op}\notag\\
&=\overline{\mathcal I}_t(Z).
\label{eq:ucb_validity_appendix}
\end{align}
Thus, on $\btheta_1\in\mathcal C_t$, we have $\mathcal I(Z,\btheta_1)\le\overline{\mathcal I}_t(Z)$ for \emph{every} feasible candidate subset, including an oracle one.

For the converse gap bound, put $\vh_t=\hat{\btheta}_t-\btheta_1$ and $q_t(Z)=\|\vV_t^{1/2}\vS_Z\vV_t^{1/2}\|_{\rm op}$. Positive semidefiniteness implies
\begin{equation}
\vS_Z\preceq q_t(Z)\vS_t,
\qquad \vS_Z\vV_t\vS_Z\preceq q_t(Z)\vS_Z,
\label{eq:leverage_matrix_bounds}
\end{equation}
since $A=\vV_t^{1/2}\vS_Z\vV_t^{1/2}\succeq0$ satisfies $A^2\preceq\|A\|_{\rm op}A$. On the coverage event $\|\vh_t\|_{\vS_t}\le\alpha_t$, these inequalities give
\begin{align}
0\le\overline{\mathcal I}_t(Z)-\mathcal I(Z,\btheta_1)
&\le 2\alpha_t\|\btheta_1\|_{\vS_Z}\sqrt{q_t(Z)}
+2\alpha_t^2q_t(Z).
\label{eq:ucb_gap_general}
\end{align}
In particular, Assumption~3 bounds the gap at the \emph{selected} subset by
\begin{equation}
0\le\overline{\mathcal I}_t(Z_{t+1})-\mathcal I(Z_{t+1},\btheta_1)
\le \frac{2c_1\sqrt{c_2}\,\alpha_t}{\sqrt w}
+\frac{2c_2\alpha_t^2}{w}
=:\epsilon_{\mathrm{ucb}}.
\label{eq:ucb_gap_selected}
\end{equation}
With the radius of Corollary~2 and $a\log(2d/\delta)=o(w)$,
\begin{equation}
\epsilon_{\mathrm{ucb}}=
\mathcal O\!\left(\sqrt{\frac{a\log(2d/\delta)}{w}}+\frac{a\log(2d/\delta)}w\right)
=\mathcal O\!\left(\sqrt{\frac{a\log(2d/\delta)}w}\right).
\label{eq:epsilon_ucb_order}
\end{equation}
Finally, suppose that the subset selection at this epoch satisfies
\begin{equation}
\overline{\mathcal I}_t(Z_{t+1})\ge
c\max_{|Z|=m}\overline{\mathcal I}_t(Z)
\label{eq:subset_approx_property_appendix}
\end{equation}
for some fixed $c\in(0,1]$. Under exact UCB maximization this is automatic with $c=1$; for greedy selection it follows only \emph{when the hypotheses of Lemma~1 hold}. If $Z^\star\in\arg\max_{|Z|=m}\mathcal I(Z,\btheta_1)$, the all-candidate UCB property yields
\[
\overline{\mathcal I}_t(Z_{t+1})
\ge c\overline{\mathcal I}_t(Z^\star)
\ge c\mathcal I^*(\btheta_1).
\]
Subtracting \eqref{eq:ucb_gap_selected} proves
$\mathcal I(Z_{t+1},\btheta_1)\ge c\mathcal I^*(\btheta_1)-\epsilon_{\rm ucb}$, as claimed.
\end{proof}

\subsection{Proof of Theorem 3}
\label{app:proof_theorem3_revised}

We derive the detection-delay bound from a positive delayed conditional drift and a blockwise first-passage estimate. Fix a detectable parameter $\btheta_1$ with $0<\mathcal I^*(\btheta_1)<\infty$. The conditions in Assumptions~1--3 and, for greedy selection, the approximation factor $c\in(0,1]$ are assumed to hold uniformly across mature post-change epochs, change points, and pre-change histories. We further require $\vS_t\succ0$ at those epochs so that the confidence-based sensing rule is applicable. We set $\delta=w^{-4}$ in the confidence radius and Lasso regularization, with $a\log(dw)=o(w)$.

For the expected-delay claim we also require the following conditional fourth-moment bound with a deterministic constant $c_3<\infty$ uniform in $t,\tau_0,w,\gamma$ and the pre-change history:
\begin{equation}
\mathbb E_{\tau_0,\btheta_1}\!\left[
\|\hat{\btheta}_{t-1}-\btheta_1\|_{\vS_{Z_t}}^4
\,\middle|\,\mathcal F_{t-w-1}\right]\le c_3,
\qquad t\ge\tau_0+w.
\label{eq:uniform_integrability_delay}
\end{equation}
This condition can either be imposed directly in a growing-dimensional regime or deduced from the Gaussian Lasso model in the bounded-information regime proved next; it is not inferred from a high-probability bound without a tail argument.

\begin{appendixlemma}[Fourth Moment of the Adaptive Lasso]
\label{lem:lasso_fourth_moment}
Suppose $m$ is fixed, $\sup_{|Z|=m}\|\vS_Z\|_{\rm op}\le L<\infty$, $\|\btheta_1\|_1\le B<\infty$, and $\sigma_{\rm sc}\ge\underline\sigma>0$, where the bounds are independent of $w$. Under the uniform full-rank and selected-action leverage conditions, the Lasso estimate satisfies \eqref{eq:uniform_integrability_delay} for $\delta=w^{-4}$ whenever $a\log(dw)=o(w)$.
\end{appendixlemma}

\begin{proof}[Proof of Lemma~\ref{lem:lasso_fourth_moment}]
Fix $t\ge\tau_0+w$ and condition on $\mathcal F_{t-w-1}$. Let $\bxi_i=\vY_{Z_i}-\bPhi_{Z_i}\btheta_1$ for $i=t-w,\ldots,t-1$, and put
\[
H_{t-1}=\frac{1}{2w}\sum_{i=t-w}^{t-1}
\|\bOmega_{Z_i}^{-1/2}\bxi_i\|_2^2.
\]
Comparison of the Lasso objective at $\hat{\btheta}_{t-1}$ and the feasible point $\btheta_1$ gives the deterministic bound
\begin{equation}
\|\hat{\btheta}_{t-1}\|_1\le
\|\btheta_1\|_1+\frac{H_{t-1}}{\lambda_{t-1}}.
\label{eq:lasso_raw_l1_bound}
\end{equation}
Conditional on $\mathcal F_{i-1}$, each squared whitened innovation has the $\chi_m^2$ distribution, regardless of the adaptive choice of $Z_i$. Conditional Jensen's inequality and iterated expectations therefore give $\mathbb E[H_{t-1}^8\mid\mathcal F_{t-w-1}]\le C_m$, with $C_m<\infty$ depending only on the fixed sensing budget.

Set $v_t=\|\hat{\btheta}_{t-1}-\btheta_1\|_{\vS_{Z_t}}^2$ and $\mathcal E_{t-1}=\{\btheta_1\in\mathcal C_{t-1}\}$. Uniform boundedness of the candidate information matrices and \eqref{eq:lasso_raw_l1_bound} imply
\begin{equation}
\mathbb E[v_t^4\mid\mathcal F_{t-w-1}]
\le C L^4\left(B^8+\lambda_{t-1}^{-8}C_m\right)
\label{eq:raw_fourth_error_bound}
\end{equation}
for a numerical constant $C$. On $\mathcal E_{t-1}$, Assumption~3 implies $v_t\le c_2\alpha_{t-1}^2/w$, whereas conditional coverage gives $\mathbb P(\mathcal E_{t-1}^c\mid\mathcal F_{t-w-1})\le\delta$. Applying conditional Cauchy--Schwarz to the complement yields
\begin{align}
\mathbb E[v_t^2\mid\mathcal F_{t-w-1}]
&\le\frac{c_2^2\alpha_{t-1}^4}{w^2}
+ C L^2\left(B^4+\lambda_{t-1}^{-4}C_m^{1/2}\right)\delta^{1/2}.
\label{eq:derived_fourth_moment}
\end{align}
For $\delta=w^{-4}$, $\alpha_{t-1}^2=\mathcal O\{a\log(dw)\}$ and $\lambda_{t-1}^{-4}=\mathcal O\{w^2/\log^2(dw)\}$. The right-hand side of \eqref{eq:derived_fourth_moment} is bounded uniformly in the conditioning history, $t$, $w$, and $\gamma$, proving \eqref{eq:uniform_integrability_delay}. No truncation of the Lasso estimate is required.
\end{proof}

\begin{appendixlemma}[Lower Bound on Delayed Conditional Drift]
\label{lem:drift_lower_bound}
For each $t\ge\tau_0+w$, under the preceding conditions,
\begin{equation}
\mathbb E_{\tau_0,\btheta_1}[\ell_t\mid\mathcal F_{t-w-1}]\ge\mu_w,
\label{eq:drift_lower_bound}
\end{equation}
where $\mu_w=c\mathcal I^*(\btheta_1)-\epsilon_{\rm ucb}-\epsilon_{\rm est}-\delta c\mathcal I^*(\btheta_1)$ and
\begin{equation}
\epsilon_{\rm est}=
\frac{c_2\alpha_{t-1}^2}{2w}+\frac{\sqrt{c_3\delta}}2
=\mathcal O\!\left(\frac{a\log(dw)}w\right).
\label{eq:mu_w_definition}
\end{equation}
Here $\epsilon_{\rm ucb}$ is the deterministic upper bound in \eqref{eq:ucb_gap_selected}, and any uniform choice of the confidence radius gives a deterministic upper bound for $\epsilon_{\rm est}$.
\end{appendixlemma}

\begin{proof}[Proof of Lemma~\ref{lem:drift_lower_bound}]
Since $Z_t$ and $\hat{\btheta}_{t-1}$ are $\mathcal F_{t-1}$-measurable, the conditional mean of the plug-in LLR under the post-change distribution is exactly
\begin{align}
\mathbb E_{\tau_0,\btheta_1}[\ell_t\mid\mathcal F_{t-1}]
&=\hat{\btheta}_{t-1}^\top\vS_{Z_t}\btheta_1
-\frac12\hat{\btheta}_{t-1}^\top\vS_{Z_t}\hat{\btheta}_{t-1}\notag\\
&=\mathcal I(Z_t,\btheta_1)
-\frac12\|\hat{\btheta}_{t-1}-\btheta_1\|_{\vS_{Z_t}}^2.
\label{eq:conditional_llr_drift_identity}
\end{align}
Let $\mathcal E_{t-1}=\{\btheta_1\in\mathcal C_{t-1}\}$. The window used to compute $\hat{\btheta}_{t-1}$ is $\{t-w,\ldots,t-1\}$ and, because $t\ge\tau_0+w$, contains only post-change observations. The martingale concentration proof of Theorem~2, conditioned on $\mathcal F_{t-w-1}$, yields
\begin{equation}
\mathbb P_{\tau_0,\btheta_1}(\mathcal E_{t-1}\mid\mathcal F_{t-w-1})\ge1-\delta.
\label{eq:window_conditional_coverage}
\end{equation}
On this event the near-oracle bound of Lemma~3 applies. Since $\mathcal I(Z_t,\btheta_1)\ge0$ outside the event,
\begin{equation}
\mathbb E_{\tau_0,\btheta_1}[\mathcal I(Z_t,\btheta_1)\mid\mathcal F_{t-w-1}]
\ge (1-\delta)c\mathcal I^*(\btheta_1)-\epsilon_{\rm ucb}.
\label{eq:expected_information_lower}
\end{equation}
Moreover, by the selected-action leverage condition in Assumption~3,
\begin{equation}
\mathbf1_{\mathcal E_{t-1}}
\|\hat{\btheta}_{t-1}-\btheta_1\|_{\vS_{Z_t}}^2
\le\frac{c_2\alpha_{t-1}^2}{w}.
\label{eq:estimation_loss_good_event}
\end{equation}
Conditional Cauchy--Schwarz and \eqref{eq:uniform_integrability_delay} bound the contribution of the complement by $\sqrt{c_3\delta}$. Therefore,
\begin{equation}
\frac12\mathbb E_{\tau_0,\btheta_1}\!\left[
\|\hat{\btheta}_{t-1}-\btheta_1\|_{\vS_{Z_t}}^2
\,\middle|\,\mathcal F_{t-w-1}\right]
\le\frac{c_2\alpha_{t-1}^2}{2w}+\frac{\sqrt{c_3\delta}}2
=:\epsilon_{\rm est}.
\label{eq:epsilon_est_bound}
\end{equation}
Combining \eqref{eq:conditional_llr_drift_identity}--\eqref{eq:epsilon_est_bound} proves \eqref{eq:drift_lower_bound}. With $\delta=w^{-4}$, $\alpha_{t-1}^2=\mathcal O\{a\log(dw)\}$, so $\mu_w=c\mathcal I^*(\btheta_1)+o(1)$ as soon as $a\log(dw)=o(w)$.
\end{proof}

\begin{appendixlemma}[Uniform Delayed Conditional Second Moment]
\label{lem:delayed_second_moment}
Under \eqref{eq:uniform_integrability_delay}, there is a deterministic $K<\infty$, independent of $t,\tau_0,w,\gamma$, such that, for $t\ge\tau_0+w$,
\begin{equation}
\mathbb E_{\tau_0,\btheta_1}[\ell_t^2\mid\mathcal F_{t-w-1}]\le K.
\label{eq:delayed_llr_second_moment}
\end{equation}
\end{appendixlemma}

\begin{proof}[Proof of Lemma~\ref{lem:delayed_second_moment}]
Write $\vh_{t-1}=\hat{\btheta}_{t-1}-\btheta_1$, $v_t=\|\vh_{t-1}\|_{\vS_{Z_t}}^2$, and $I_t=\mathcal I(Z_t,\btheta_1)\le\mathcal I^*(\btheta_1)$. The conditional Gaussian decomposition is
\[
\ell_t=I_t-\tfrac12v_t+\eta_t,
\qquad
\mathbb E[\eta_t\mid\mathcal F_{t-1}]=0,
\quad
\mathbb E[\eta_t^2\mid\mathcal F_{t-1}]
=\|\hat{\btheta}_{t-1}\|_{\vS_{Z_t}}^2.
\]
Using $(x-y)^2\le2x^2+2y^2$ and
$\|\hat{\btheta}_{t-1}\|_{\vS_{Z_t}}^2\le4I_t+2v_t$ gives
\begin{equation}
\mathbb E[\ell_t^2\mid\mathcal F_{t-1}]
\le2\{\mathcal I^*(\btheta_1)\}^2+\tfrac12v_t^2
+4\mathcal I^*(\btheta_1)+2v_t.
\label{eq:llr_second_decomposition}
\end{equation}
Take conditional expectations given $\mathcal F_{t-w-1}$, invoke \eqref{eq:uniform_integrability_delay}, and use conditional Cauchy--Schwarz, $\mathbb E[v_t\mid\mathcal F_{t-w-1}]\le\sqrt{c_3}$. Thus \eqref{eq:delayed_llr_second_moment} holds with
$K=2\{\mathcal I^*(\btheta_1)\}^2+\tfrac12c_3+4\mathcal I^*(\btheta_1)+2\sqrt{c_3}$.
\end{proof}

\begin{appendixlemma}[Blockwise Hitting-Time Bound]
\label{lem:blockwise_hitting_time}
Let $\{X_t\}$ be an adapted sequence with a filtration $\{\mathcal F_t\}$ and let $w$ be a positive integer. Suppose that for all $t\ge\tau_0+w$,
\[
\mathbb E[X_t\mid\mathcal F_{t-w-1}]\ge\mu>0,
\qquad
\mathbb E[X_t^2\mid\mathcal F_{t-w-1}]\le K
\]
almost surely, uniformly in the conditioning histories and change points. For $b>0$, extend the predictable sensing and testing recursion counterfactually after an alarm and define $T_b$ as the first threshold crossing at or after $\tau_0$ on the extended path; a threshold crossing before $\tau_0$ contributes zero to the actual Lorden delay. For any integer $L$ satisfying $\mu L>b$, put
\begin{equation}
q_{w,L,b}=\frac{K(2w+1)L}{(\mu L-b)^2}.
\label{eq:block_failure_bound}
\end{equation}
If $q_{w,L,b}<1$, then
\begin{equation}
\operatorname*{ess\,sup}
\mathbb E[(T_b-\tau_0+1)^+\mid\mathcal F_{\tau_0-1}]
\le\frac{w+L}{1-q_{w,L,b}}.
\label{eq:block_expected_delay_bound}
\end{equation}
\end{appendixlemma}

\begin{proof}[Proof of Lemma~\ref{lem:blockwise_hitting_time}]
Fix a deterministic block start $s\ge\tau_0$, condition on $\mathcal F_{s-1}$ and use the last $L$ observations of the block $\{s,\ldots,s+w+L-1\}$. For $j=s+w,\ldots,s+w+L-1$, put
\[
D_j=X_j-\mathbb E[X_j\mid\mathcal F_{j-w-1}].
\]
Since $\mathcal F_{s-1}\subseteq\mathcal F_{j-w-1}$, the assumptions give
$\mathbb E[D_j\mid\mathcal F_{j-w-1}]=0$ and
$\mathbb E[D_j^2\mid\mathcal F_{s-1}]\le K$.
For $v-u\ge w+1$, $D_u$ is $\mathcal F_{v-w-1}$-measurable, so
$\mathbb E[D_uD_v\mid\mathcal F_{s-1}]=0$.
There are at most $(2w+1)L$ pairs with $|u-v|\le w$. Conditional Cauchy--Schwarz therefore shows
\begin{equation}
\mathbb E\!\left[\left(\sum_{j=s+w}^{s+w+L-1}D_j\right)^2
\,\middle|\,\mathcal F_{s-1}\right]
\le K(2w+1)L.
\label{eq:block_delay_variance}
\end{equation}
The sum of the corresponding delayed conditional means is at least $\mu L$. Conditional Chebyshev thus yields
\begin{equation}
\mathbb P\!\left(
\sum_{j=s+w}^{s+w+L-1}X_j\le b
\,\middle|\,\mathcal F_{s-1}\right)
\le q_{w,L,b}.
\label{eq:block_no_cross_probability}
\end{equation}
The reflected statistic at the end of the block is at least the sum of the last $L$ increments. Hence survival through this block implies the event in \eqref{eq:block_no_cross_probability}. Partition the time after $\tau_0$ into consecutive deterministic blocks of length $B=w+L$. Iterated conditioning gives
\[
\mathbb P(T_b-\tau_0+1>kB\mid\mathcal F_{\tau_0-1})\le q_{w,L,b}^{\,k},
\qquad k\ge0.
\]
Summing this geometric tail in blocks proves \eqref{eq:block_expected_delay_bound}. The argument uses only the delayed conditional mean, delayed second moment, and predictability of the increments; no independence or overshoot assumption is needed.
\end{proof}

\begin{proof}[Proof of Theorem~3]
Set $b=\log\gamma$ and keep an arbitrary change point and pre-change history fixed. The actual detection delay is no greater than the delay to the first post-change crossing on the counterfactually continued recursion. By Lemmas~\ref{lem:drift_lower_bound} and~\ref{lem:delayed_second_moment}, the post-change plug-in increments satisfy the hypotheses of Lemma~\ref{lem:blockwise_hitting_time} with
\begin{equation}
\mu_w=c\mathcal I^*(\btheta_1)-\epsilon_{\rm ucb}
-\epsilon_{\rm est}-w^{-4}c\mathcal I^*(\btheta_1)
=c\mathcal I^*(\btheta_1)+o(1)>0.
\label{eq:mu_w_limit}
\end{equation}
Let $\varepsilon_b=(w/b)^{1/4}$ and choose
\begin{equation}
L_b=\left\lceil\frac{(1+\varepsilon_b)(b+1)}{\mu_w}\right\rceil.
\label{eq:block_length_choice}
\end{equation}
Since $w=o(b)$ and $\mu_w$ converges to a strictly positive constant, $L_b=\mathcal O(b)$ and
\begin{equation}
q_{w,L_b,b}
=\mathcal O\!\left(\frac{w}{\varepsilon_b^2 b}\right)
=\mathcal O\!\left(\sqrt{\frac wb}\right)=o(1).
\label{eq:block_failure_asymptotic}
\end{equation}
Using \eqref{eq:block_expected_delay_bound} and taking the supremum over the change point and essential supremum over pre-change histories gives
\begin{align}
J_{\btheta_1}^{\rm FAST}(Z,\tau_1)
&\le\frac{w+L_b}{1-q_{w,L_b,b}}
\notag\\
&=\frac{b}{\mu_w}\{1+o(1)\}
=\frac{\log\gamma}{c\mathcal I^*(\btheta_1)}\{1+o(1)\}.
\label{eq:windowed_wald_bound}
\end{align}
This proves Theorem~3 under the explicit conditions stated above. Lemma~2 continues to give $\mathbb E_\infty[\tau_1]\ge\gamma$, including any predictable rank-completion stage. Combining the established upper bound with Theorem~1 proves Corollary~3. If the UCB subset is maximized exactly, $c=1$ and the asymptotic ratio is one because FAST is itself an admissible procedure.
\end{proof}

The proof uses only the delayed conditional moment bounds of the plug-in increments; neither independence nor an i.i.d. overshoot approximation is needed. Lemma~\ref{lem:lasso_fourth_moment} provides the fourth-moment bound in the fixed-budget bounded-information setting, while \eqref{eq:uniform_integrability_delay} remains an explicit assumption in more general regimes. Likewise, the full-rank and approximation properties invoked above are conditions of the delay result, rather than automatic consequences of the UCB or greedy sensing rule.

\section{Detection Performance Comparison}
\label{app:numerical}
\begin{table}[!h]
\centering
\caption{Detection performance comparison in terms of ADD under different scale regimes for Pattern I. Numbers in parentheses report the standard deviations of detection delays across replications.}
\label{tab:add_ab1}
\scalebox{1}{
\begin{tabular}{ccccccccc}
\toprule
Scale & Shift & Oracle & FAST & FUCB & SRAS & TS & CMAB \\
\midrule
% Small
\multirow{10}{*}{\rotatebox{270}{Small}}
&0&200(194)&200(185)&200(186)&200(189)&200(197)&200(194)\\
&0.2&61.4(65.1)&92.1(89.2)&112(111)&91.3(87.2)&130(133)&167(172)\\
&0.4&19.3(15.0)&32.4(27.1)&42.3(39.6)&40.7(33.9)&51.2(45.9)&120(147)\\
&0.6&10.7(6.45)&14.9(8.94)&17.9(13.4)&25.9(17.0)&23.3(19.0)&81.5(114)\\
&0.8&7.50(3.24)&9.37(5.04)&10.8(6.19)&16.6(8.75)&12.9(10.0)&57.9(115)\\
&1.0&5.90(2.40)&7.43(3.40)&7.78(4.06)&12.6(6.12)&8.45(5.44)&58.0(130)\\
&1.2&5.02(1.91)&6.25(2.78)&6.14(2.78)&8.94(4.06)&5.72(3.52)&46.8(114)\\
&1.4&4.35(1.58)&5.48(2.56)&5.37(2.24)&5.42(2.42)&4.22(2.47)&51.9(132)\\
&1.6&3.85(1.33)&4.81(2.20)&4.64(2.04)&4.89(2.22)&3.60(2.23)&38.5(106)\\
&1.8&3.54(1.18)&4.26(1.92)&4.20(1.81)&4.44(1.91)&2.95(1.81)&50.3(132)\\
\midrule
% Medium
\multirow{10}{*}{\rotatebox{270}{Medium}}
&0&200(199)&200(195)&200(188)&200(184)&200(187)&200(196)\\
&0.2&53.2(52.4)&123(123)&147(145)&130(132)&163(161)&167(170)\\
&0.4&22.7(18.5)&41.3(32.5)&67.2(61.6)&60.6(52.9)&109(104)&139(154)\\
&0.6&13.1(8.49)&21.0(12.3)&32.1(25.3)&38.1(27.9)&72.0(65.5)&106(131)\\
&0.8&9.30(5.23)&13.9(7.61)&19.1(12.9)&23.6(14.5)&42.8(36.8)&86.1(133)\\
&1.0&7.54(3.70)&10.8(5.13)&12.4(6.82)&18.1(9.70)&26.5(23.1)&75.1(132)\\
&1.2&6.22(2.86)&8.83(4.40)&9.65(4.82)&12.1(6.21)&16.2(12.9)&63.6(122)\\
&1.4&5.27(2.44)&7.89(3.79)&8.17(3.76)&6.92(3.15)&10.4(8.00)&56.7(121)\\
&1.6&4.77(2.08)&7.05(3.83)&6.91(3.26)&6.04(2.67)&7.25(5.42)&57.8(132)\\
&1.8&4.25(1.73)&6.30(3.27)&6.18(2.86)&5.41(2.36)&5.66(4.07)&60.3(132)\\
\midrule
% Large
\multirow{10}{*}{\rotatebox{270}{Large}}
&0&200(200)&200(189)&200(183)&200(185)&200(178)&200(186)\\
&0.2&34.4(36.3)&123(120)&159(157)&186(173)&161(151)&184(173)\\
&0.4&13.5(10.3)&43.2(32.8)&82.6(75.1)&163(163)&161(159)&170(168)\\
&0.6&9.20(5.87)&21.0(12.8)&34.5(26.8)&145(143)&130(130)&134(152)\\
&0.8&6.64(4.09)&14.6(8.30)&19.3(12.6)&104(103)&102(96.1)&124(164)\\
&1.0&5.47(3.10)&11.4(5.76)&12.6(6.24)&81.7(73.5)&83.1(82.2)&103(147)\\
&1.2&4.58(2.55)&9.83(5.36)&10.0(4.87)&63.9(57.2)&65.9(62.5)&91.6(157)\\
&1.4&4.09(2.10)&8.62(5.25)&8.39(4.40)&48.5(39.0)&53.4(50.6)&81.7(141)\\
&1.6&3.53(1.84)&7.30(4.52)&7.40(3.68)&36.7(27.5)&38.5(34.3)&80.6(148)\\
&1.8&3.34(1.55)&7.13(4.31)&6.51(3.53)&29.8(21.6)&28.1(26.0)&82.8(155)\\
\bottomrule
\end{tabular}}
\end{table}

\clearpage
\begin{table}[!h]
\centering
\caption{Detection performance comparison in terms of ADD under different scale regimes for Pattern II. Numbers in parentheses report the standard deviations of detection delays across replications.}
\label{tab:add_ab2}
\scalebox{1}{
\begin{tabular}{cccccccc}
\toprule
Scale & Shift & Oracle & FAST & FUCB & SRAS & TS & CMAB \\
\midrule
% Small
\multirow{10}{*}{\rotatebox{270}{Small}}
&0&200(195)&200(193)&200(188)&200(192)&200(198)&200(198)\\
&0.2&119(116)&163(157)&158(151)&172(179)&197(188)&195(188)\\
&0.4&49.8(44.7)&94.4(90.2)&112(110)&116(173)&136(128)&171(171)\\
&0.6&23.8(18.1)&48.2(44.2)&67.8(63.3)&79.4(121)&95.9(89.3)&154(155)\\
&0.8&15.3(9.15)&28.9(23.8)&39.6(34.1)&50.9(72.1)&69.8(67.3)&127(136)\\
&1.0&11.5(6.18)&18.4(12.5)&25.1(20.7)&35.7(49.5)&46.7(40.3)&101(121)\\
&1.2&8.94(3.85)&13.3(7.58)&17.9(12.7)&23.4(24.8)&32.5(26.9)&78.9(110)\\
&1.4&7.61(3.26)&10.7(5.60)&13.2(8.34)&13.5(14.9)&23.5(18.7)&74.6(117)\\
&1.6&6.59(2.61)&9.24(4.63)&10.3(5.47)&11.6(9.99)&16.7(12.3)&64.8(103)\\
&1.8&5.64(2.16)&8.23(3.69)&8.89(4.54)&9.55(7.69)&14.0(9.40)&56.9(105)\\
\midrule
% Medium
\multirow{10}{*}{\rotatebox{270}{Medium}}
&0&200(200)&200(197)&200(188)&200(189)&200(186)&200(195)\\
&0.2&101(105)&153(165)&178(185)&184(163)&180(170)&202(199)\\
&0.4&40.7(37.9)&85.6(82.0)&130(130)&172(124)&166(161)&189(183)\\
&0.6&21.6(16.9)&45.7(40.6)&86.6(82.6)&128(75.6)&146(140)&161(162)\\
&0.8&14.2(8.34)&28.1(21.4)&52.3(47.6)&82.5(45.0)&115(111)&158(160)\\
&1.0&11.1(5.89)&20.4(15.0)&35.9(29.0)&60.1(30.9)&98.0(90.4)&131(145)\\
&1.2&9.19(4.24)&16.7(11.1)&23.3(16.7)&36.4(17.7)&85.5(78.7)&120(143)\\
&1.4&8.13(3.65)&13.5(8.87)&17.4(10.9)&21.6(14.9)&70.4(62.8)&99.0(129)\\
&1.6&7.28(3.14)&11.9(7.34)&14.1(8.23)&16.5(9.99)&56.6(47.3)&93.9(127)\\
&1.8&6.58(2.83)&10.6(6.23)&11.3(5.82)&13.5(7.69)&48.2(41.1)&88.1(133)\\
\midrule
% Large
\multirow{10}{*}{\rotatebox{270}{Large}}
&0&200(201)&200(201)&200(191)&200(184)&200(172)&200(194)\\
&0.2&99.9(97.6)&181(180)&180(178)&183(177)&189(178)&204(195)\\
&0.4&44.8(39.4)&131(129)&155(148)&182(182)&167(164)&200(198)\\
&0.6&25.6(17.8)&70.3(60.2)&119(113)&153(154)&155(145)&183(173)\\
&0.8&17.1(9.38)&41.8(32.7)&76.9(70.0)&143(147)&140(139)&186(177)\\
&1.0&13.2(6.38)&29.9(21.0)&53.4(48.0)&109(104)&127(124)&171(169)\\
&1.2&10.9(4.92)&21.8(13.4)&36.3(29.6)&97.7(92.2)&113(99.6)&150(157)\\
&1.4&9.53(4.08)&18.7(12.4)&26.9(18.2)&74.2(66.0)&101(94.4)&142(155)\\
&1.6&8.41(3.62)&16.4(12.3)&20.8(12.4)&59.9(52.2)&91.8(86.4)&125(146)\\
&1.8&7.34(3.06)&15.0(11.5)&17.4(9.30)&47.4(38.2)&88.5(82.7)&119(144)\\
\bottomrule
\end{tabular}}
\end{table}

\bibliographystyle{IEEEtranN}
\bibliography{references}

\end{document}